\documentclass[reprint,superscriptaddress,amsmath,amssymb,aps]{revtex4-1}

\usepackage{graphicx}
\usepackage{dcolumn}
\usepackage{bm}
\usepackage{subfigure}
\usepackage{natbib} 
\usepackage{multirow}
\usepackage{soul}
\usepackage{float}
\usepackage[normalem]{ulem}
\usepackage[usenames,dvipsnames]{color}
\usepackage{appendix}
\usepackage{mathrsfs}
\usepackage[colorlinks=true,linkcolor=blue,citecolor=blue,urlcolor=blue]{hyperref}

\newcommand{\braket}[2] { \langle #1 | #2 \rangle }

\def\QE{\textsc{Quantum ESPRESSO}\,}

\newcommand{\editor}[2]{%
  \expandafter\newcommand\csname #1note\endcsname[1]{%
    \textcolor{#2}{(\textbf{#1:} ##1)}}%
  \expandafter\newcommand\csname #1\endcsname[1]{%
    \textcolor{#2}{##1}}%
  \expandafter\newcommand\csname #1cancel\endcsname[1]{%
    \textcolor{#2}{\sout{##1}}}%
  \expandafter\newcommand\csname #1change\endcsname[2]{%
    \textcolor{#2}{\sout{##1} ##2}}%
  \newenvironment{#1text}{\color{#2}}{\color{black}}
}

\begin{document}

\title{Importance of intersite Hubbard interactions in $\beta$-MnO$_2$: A first-principles DFT+$U$+$V$ study}

\author{Ruchika Mahajan}
\affiliation{School of Basic Sciences, Indian Institute of Technology Mandi, Himachal Pradesh 175075, India}

\author{Iurii Timrov}\email[]{ iurii.timrov@epfl.ch}
\affiliation{Theory and Simulation of Materials (THEOS), and National Centre for Computational Design and Discovery of Novel Materials (MARVEL), \'Ecole Polytechnique F\'ed\'erale de Lausanne (EPFL), CH-1015 Lausanne, Switzerland}

\author{Nicola Marzari}
\affiliation{Theory and Simulation of Materials (THEOS), and National Centre for Computational Design and Discovery of Novel Materials (MARVEL), \'Ecole Polytechnique F\'ed\'erale de Lausanne (EPFL), CH-1015 Lausanne, Switzerland}

\author{Arti Kashyap}\email[]{ arti@iitmandi.ac.in}
\affiliation{School of Basic Sciences, Indian Institute of Technology Mandi,
Himachal Pradesh 175075, India}

\begin{abstract} 
We present a first-principles investigation of the structural, electronic, and magnetic properties of pyrolusite ($\beta$-MnO$_2$) using conventional and extended Hubbard-corrected density-functional theory (DFT+$U$ and DFT+$U$+$V$). The onsite $U$ and intersite $V$ Hubbard parameters are computed using linear-response theory in the framework of density-functional perturbation theory. We show that while the inclusion of the onsite $U$ is crucial to describe the localized nature of the Mn($3d$) states, the intersite $V$ is key to capture accurately the strong hybridization between neighboring Mn($3d$) and O($2p$) states. In this framework, we stabilize the simplified collinear antiferromagnetic (AFM) ordering (suggested by the Goodenough-Kanamori rule) that is commonly used as an approximation to the experimentally-observed noncollinear screw-type spiral magnetic ordering. A detailed investigation of the ferromagnetic and of other three collinear AFM spin configurations is also presented. The findings from Hubbard-corrected DFT are discussed using two kinds of Hubbard manifolds -- nonorthogonalized and orthogonalized atomic orbitals -- showing that special attention must be given to the choice of the Hubbard projectors, with orthogonalized manifolds providing more accurate results than nonorthogonalized ones within DFT+$U$+$V$. This paper paves the way for future studies of complex transition-metal compounds containing strongly localized electrons in the presence of pronounced covalent interactions. 
\end{abstract}

\date{\today} 

\maketitle

\section{Introduction}
\label{sec:intro}

Manganese oxides have been studied for many decades because they are attractive not only from the fundamental point of view but also due to their importance for a variety of technological applications owing to their low cost, low toxicity, and high chemical stability~\cite{Ghosh:2020}. In particular, pyrolusite ($\beta$-MnO$_2$) is the most stable and abundant polymorph of the MnO$_2$ family, and it is used as a catalyst~\cite{Yang:2006, Lima:2007}, cathode in alkaline batteries~\cite{Seo:2018, Zhang:2017} and Li-ion batteries~\cite{Thackeray:1983, wang2013beta, yao2018cations} and Li-O$_2$ batteries~\cite{Thapa:2011}, electrode in supercapacitors~\cite{Zang:2011}, to name a few applications. Experimentally, $\beta$-MnO$_2$ is a small-band-gap semiconductor ($0.26-0.28$~eV~\cite{Chevillot:1959, yu2008electronic, Islam:2005, druilhe1967electron}), which exists in a tetragonal ($P4_2/mnm$) rutile structure, and it undergoes a paramagnetic to noncollinear helical (screw-type spiral) antiferromagnetic transition at $T_N = 92$~K~\cite{yoshimori1959new, sato2000transport, sato2001magnetic, F2006Enhanced}. According to Yoshimori~\cite{yoshimori1959new}, the pitch of this spiral is exactly $7c/2$, i.e., the spins lie on the $ab$ plane and rotate by $129^\circ$ in the next-adjacent layer along the $c$ axis for a period of 7 unit cells. However, more recent and more refined measurements~\cite{regulski2003incommensurate} reported that the pitch is about 4$\%$ shorter than $7c/2$, meaning that $\beta$-MnO$_2$ has an incommensurate magnetic ordering.

From the computational point of view, accurate first-principles modeling of $\beta$-MnO$_2$ with its very complex spin configuration and intricate electronic interactions (due to the partially filled $3d$-shell of Mn atoms) remains a major challenge. Density-functional theory (DFT)~\cite{Hohenberg:1964,Kohn:1965} simulations in the local spin-density approximation (LSDA) or spin-polarized generalized-gradient approximations ($\sigma$-GGA) for the exchange-correlation (xc) functional -- which is a workhorse of the materials science -- are unable to provide satisfactory results for many transition-metal compounds including $\beta$-MnO$_2$~\cite{franchini2007ground}. This is due to the large self-interaction errors (SIE)~\cite{Perdew:1981, MoriSanchez:2006} especially which are significant for localized $d$ and $f$ electrons. For this reason, more accurate approaches beyond standard DFT are generally applied. Particularly for $\beta$-MnO$_2$ many approaches have been used, with the majority of the studies based on Hubbard-corrected DFT (so-called DFT+$U$~\cite{anisimov:1991, Liechtenstein:1995, dudarev1998electron})~\cite{franchini2007ground, tompsett2012importance, Cockayne:2012, wang2013beta, dawson2015first, lim2016improved, Wang:2016, kitchaev2016energetics, xu2017role, chepkoech2018first} and DFT with hybrid xc functionals (PBE0~\cite{Adamo:1999} and HSE06~\cite{Heyd:2003, Heyd:2006})~\cite{franchini2007ground, tompsett2012importance, wang2013beta, kitchaev2016energetics, xu2017role}. In the former approach, the Hubbard $U$ correction is applied selectively only to the partially filled $d$ states to alleviate SIE for these states~\cite{Kulik:2008}, while all other states are treated at the level of LSDA or $\sigma$-GGA, while in hybrid functionals a fraction of exact exchange (Fock) is added (25$\%$ in the case of PBE0 and HSE06) and the remainder of exchange is treated at the $\sigma$-GGA level, together with 100$\%$ of the $\sigma$-GGA correlation. On one hand, the advantage of DFT+$U$ is that it often greatly improves standard DFT with only marginally more computationally expensive calculations; still, the value of the Hubbard $U$ parameter has to be determined in some satisfactory way. On the other hand, hybrid functionals are more accurate than standard DFT, but they are computationally more expensive than DFT or DFT+$U$ calculations. Furthermore, for hybrid functionals, quite often the required fraction of exact exchange must be tuned in solids to reach better agreement with experiments (although there are \textit{ab initio} methods to determine the amount of exact exchange needed for a system of interest ~\cite{Skone:2014, Skone:2016, Bischoff:2019, Kronik:2012, Wing:2021, Lorke:2020}). $\beta$-MnO$_2$ has also been studied using other methods: Hartree-Fock~\cite{mackrodt1997first}, tight-binding~\cite{Zhuang:2001}, DFT plus dynamical mean field theory (DMFT)~\cite{yu2008electronic}, DFT with the SCAN meta-GGA functional~\cite{kitchaev2016energetics, gautam2018evaluating}, and SCAN+$U$~\cite{gautam2018evaluating}.

Modelling the incommensurate magnetic pattern of $\beta$-MnO$_2$ is computationally demanding as the simulation cell is very large and beyond reach for some of the aforementioned methods (e.g., hybrid functionals). For this reason, the vast majority of the first-principles studies of this material are done using simplified collinear magnetic ordering~\cite{mackrodt1997first, franchini2007ground, tompsett2012importance, wang2013beta, dawson2015first, noda2016momentum, Wang:2016, kitchaev2016energetics, xu2017role, chepkoech2018first, gautam2018evaluating}, with only a few studies done using a noncollinear helical magnetic ordering with the $7c/2$ pitch~\cite{Zhuang:2001, yu2008electronic, tompsett2012importance, lim2016improved}. The use of the simplified collinear magnetic ordering is motivated by the fact that many studies need to go beyond the standard electronic-structure analysis and investigate other properties of $\beta$-MnO$_2$; these include thermoelectric~\cite{chepkoech2018first} and thermochemical~\cite{gautam2018evaluating} properties, the formation of oxygen vacancies~\cite{dawson2015first, xu2017role}, intercalation voltages and kinetics of Li diffusion in MnO$_2$-based cathodes in Li-ion batteries~\cite{wang2013beta}.

For the collinear case, all studies consider typically two magnetic orderings, namely a ferromagnetic (FM) one and an antiferromagnetic (AFM) ordering where the latter corresponds to opposing spins on the center and corner Mn sites in the unit cell (what will be called A1-AFM in this paper). The choice of this specific collinear AFM ordering is based on the Goodenough-Kanamori rule~\cite{Goodenough:1955, Kanamori:1959}, which suggests that this spin configuration should be the most preferable one in $\beta$-MnO$_2$~\cite{noda2016momentum}. In the DFT+$U$ and hybrid functional studies mentioned above, the total energies of the FM and A1-AFM orderings are compared: All hybrid functional studies predict the A1-AFM ordering to be lower in energy than the FM one, while DFT+$U$ studies give different results depending on the value of the Hubbard $U$ parameter used. Commonly, in $\beta$-MnO$_2$ studies the $U$ parameter is determined empirically, such that DFT+$U$ reproduces well some experimental property of interest (e.g. the band gap or reaction enthalpy)~\cite{franchini2007ground, yu2008electronic, Cockayne:2012, wang2013beta, lim2016improved, noda2016momentum, kitchaev2016energetics, chepkoech2018first, gautam2018evaluating}; very rarely it is computed from first principles~\cite{tompsett2012importance}. Most importantly, when comparing various $U$ values from different works it is often forgotten to check which Hubbard manifold (projectors) are used -- as was shown in Ref.~\cite{Kulik:2008, Wang:2016} and as will be discussed in detail in this paper, the values of $U$ are not universal, but dependent on the type of the Hubbard projectors that are used. Hence, it is crucial to keep consistency between Hubbard parameters and Hubbard projectors, and keep this in mind when comparing DFT+$U$ results from different papers~\cite{Note:Rappe_vs_Kresse}. The reported empirical $U$ values for $\beta$-MnO$_2$ span a very wide range going from 1 to 7~eV; generally it was found that, using projector-augmented-wave (PAW) Hubbard projectors as implemented in VASP~\cite{Rohrbach:2003}, for smaller values of $U$ the A1-AFM ordering is lower in energy than FM, while for larger values of $U$ the FM ordering is lower in energy than A1-AFM~\cite{franchini2007ground, noda2016momentum, Wang:2016}. However, it is always assumed that A1-AFM is the only AFM ordering that should be considered among other AFM orderings, with the only exception being Ref.~\cite{noda2016momentum} that considered also another type of the AFM ordering (what will be called A2-AFM in this paper) at the DFT+$U$ level. A detailed investigation of whether A1-AFM is indeed the most energetically favorable spin configuration with respect to various other collinear AFM orderings for $\beta$-MnO$_2$ and how this depends on the type of Hubbard projectors is currently missing in the literature; this point will be addressed in this paper.

In the majority of Hubbard-corrected DFT studies of $\beta$-MnO$_2$, a simplified rotationally invariant formulation of DFT+$U$~\cite{dudarev1998electron} is used with an effective interaction parameter $U_\mathrm{eff} = U-J$, where $U$ representing a screened onsite Coulomb repulsion and $J$ is the Hund's exchange parameter. It has been shown by Tompsett \textit{et al.}~\cite{tompsett2012importance} that at this level of theory with $U_\mathrm{eff} = 5.5$~eV (determined from first principles using constrained DFT~\cite{Madsen:2005}) within the linearized augmented plane wave (LAPW) method, the FM spin configuration is lower in energy than A1-AFM. Instead, when Hund's $J=1.2$~eV is treated explicitly on the same footing as $U=6.7$~eV (instead of one effective parameter $U_\mathrm{eff}$), A1-AFM is lower in energy than FM. However, a more recent paper by Wang \textit{et al.}~\cite{Wang:2016} has shown that even with one effective parameter $U_\mathrm{eff} = 5.5$~eV it is possible to have A1-AFM lower in energy than FM for certain types of Hubbard projectors (different from those of the LAPW method), thus underlying the crucial role played not only by the value of the Hubbard parameters but also by the Hubbard projectors~\cite{Kick:2019}. Therefore, it would be important to further investigate the effect of the explicit account of Hund's $J$ (like in Ref.~\cite{tompsett2012importance}) depending on different types of Hubbard projectors~\cite{Timrov:2020b}. Furthermore, another important aspect that has been disregarded so far is the investigation of the importance of the intersite Hubbard interactions between Mn($3d$) and O($2p$) states, that are known to be strongly hybridized in the valence region~\cite{mackrodt1997first}. In fact, it has been pointed out by Yu \textit{et al.}~\cite{yu2008electronic} that it is very important to take into account the Mn($3d$)--O($2p$) hybridizations when constructing Wannier functions for DFT+DMFT for an accurate description of properties of $\beta$-MnO$_2$. Therefore, this aspect will also be investigated in detail here.

In this paper, we present a detailed investigation of the structural, electronic, and magnetic properties of $\beta$-MnO$_2$ using DFT+$U$ and its extension DFT+$U$+$V$~\cite{campo2010extended}, where $V$ is the intersite Hubbard parameter. The interaction parameters $U$ and $V$ are computed self-consistently from first-principles using linear-response theory~\cite{cococcioni2005linear} reformulated in terms of density-functional perturbation theory (DFPT)~\cite{timrov2018hubbard, Timrov:2021}. The study is carried out using two types of Hubbard projectors, namely nonorthogonalized atomic orbitals (NAO) and orthogonalized atomic orbitals (OAO)~\cite{Timrov:2020b}; in OAO the orthogonalization insures that Hubbard corrections are applied only once to the respective Hubbard manifolds. We investigate five collinear magnetic orderings of $\beta$-MnO$_2$: FM, A1-AFM, A2-AFM, C-AFM, and G-AFM. Based on this framework, we show the crucial role played by the intersite Hubbard interactions due to the strong covalent Mn($3d$)-O($2p$) hybridizations and we highlight the importance of a careful choice of the type of Hubbard projectors and especially of having Hubbard parameters that are consistent with those projectors. Our detailed analysis of ground-state properties shows that overall A1-AFM is the most energetically favorable and the most accurate collinear (simplified) representation of the real noncollinear helical magnetic ordering of $\beta$-MnO$_2$, provided that the DFT+$U$+$V$ formulation with first-principles $U$ and $V$ and OAO Hubbard projectors are used, since they properly describe the delicate interplay between structural, electronic, and magnetic degrees of freedom in this material.     

The paper is organized as follows. Section~\ref{sec:methods} presents the basics of DFT+$U$ and DFT+$U$+$V$ approaches and the DFPT approach for computing $U$ and $V$; Sec.~\ref{sec:technical_details} contains technical details of our calculations; in Sec.~\ref{sec:Results_and_Discussion} we present our findings for the structural, electronic, and magnetic properties of $\beta$-MnO$_2$ for five types of collinear magnetic orderings; and in Sec.~\ref{sec:Conclusions} we give our conclusions.

\section{Computational method}
\label{sec:methods}

In this section we briefly discuss the basics of the DFT+$U$+$V$ approach~\cite{campo2010extended, Himmetoglu:2014} and of the DFPT approach for computing Hubbard parameters~\cite{timrov2018hubbard, Timrov:2021}. All equations in this subsection can be easily reduced to the DFT+$U$ case by simply setting $V=0$. For the sake of simplicity, the formalism is presented in the framework of norm-conserving (NC) pseudopotentials (PPs) in the collinear spin-polarized case. The generalization to the ultrasoft (US) PPs and the projector augmented wave (PAW) method can be found in Ref.~\cite{Timrov:2021}. Hartree atomic units are used.

\subsection{DFT+$U$+$V$}

In DFT+$U$+$V$, a correction term is added to the approximate DFT energy functional~\cite{campo2010extended}: 
\begin{equation}
E_{\mathrm{DFT}+U+V} = E_{\mathrm{DFT}} + E_{U+V} ,
\label{eq:Edft_plus_u}
\end{equation}
where $E_{\mathrm{DFT}}$ is the approximate DFT energy (constructed, e.g., within  LSDA or $\sigma$-GGA), and $E_{U+V}$ contains the additional Hubbard term. At variance with the DFT+$U$ approach, containing only onsite interactions scaled by $U$, DFT+$U$+$V$ contains also intersite interactions between an atom and its surrounding ligands scaled by $V$. In the case of $\beta$-MnO$_2$, the onsite $U$ correction is needed for the Mn($3d$) states, while the intersite $V$ correction is expected to be relevant to describe Mn($3d$)--O($2p$) interactions~\cite{Kulik:2011}. In the simplified rotationally-invariant formulation~\cite{dudarev1998electron}, the extended Hubbard term reads:
\begin{eqnarray}
E_{U+V} & = & \frac{1}{2} \sum_I \sum_{\sigma m m'} 
U^I \left( \delta_{m m'} - n^{II \sigma}_{m m'} \right) n^{II \sigma}_{m' m} \nonumber \\
& & - \frac{1}{2} \sum_{I} \sum_{J (J \ne I)}^* \sum_{\sigma m m'} V^{I J} 
n^{I J \sigma}_{m m'} n^{J I \sigma}_{m' m} \,,
\label{eq:Edftu}
\end{eqnarray}
where $I$ and $J$ are atomic site indices, $m$ and $m'$ are the magnetic quantum numbers associated with a specific angular momentum [$l=2$ for Mn($3d$) and $l=1$ for O($2p$)], $U^I$ and $V^{I J}$ are the effective onsite and intersite Hubbard parameters (we dropped the subscript ``eff'' for simplicity), and the star in the sum denotes that for each atom $I$, the index $J$ covers all its neighbors up to a given distance (or up to a given shell). 

The generalized occupation matrices $n^{I J \sigma}_{m m'}$ are based on a projection of the Kohn-Sham (KS) states on localized orbitals $\phi^{I}_{m}(\mathbf{r})$ of neighbor atoms: 
\begin{equation}
n^{I J \sigma}_{m m'} = \sum_{v,\mathbf{k}} f^\sigma_{v,\mathbf{k}}
\braket{\psi^\sigma_{v,\mathbf{k}}}{\phi^{J}_{m'}} \braket{\phi^{I}_{m}}{\psi^\sigma_{v,\mathbf{k}}} \,, 
\label{eq:occ_matrix_0}
\end{equation}
where $v$ and $\sigma$ represent, respectively, the band and spin labels of the KS wavefunctions $\psi^\sigma_{v,\mathbf{k}}(\mathbf{r})$, $\mathbf{k}$ indicate points in the first Brillouin zone (BZ), $f^\sigma_{v,\mathbf{k}}$ are the occupations of the KS states, and $\phi^I_{m}(\mathbf{r}) \equiv \phi^{\gamma(I)}_{m}(\mathbf{r} - \mathbf{R}_I)$ are localized orbitals centered on the $I$th atom of type $\gamma(I)$ at the position $\mathbf{R}_I$. It is convenient to establish a short-hand notation for the onsite occupation matrix: $n^{I\sigma}_{m m'} \equiv n^{II\sigma}_{m m'}$, which is used in the standard DFT+$U$ approach that corresponds to the first line of Eq.~\eqref{eq:Edftu}. The two terms in Eq.~\eqref{eq:Edftu} (i.e., proportional to the onsite $U^{I}$ and intersite $V^{IJ}$ couplings) counteract each other: the onsite term favors localization on atomic sites (thus suppressing hybridization with neighbors), while the intersite term favors hybridized states with components on neighbor atoms. More details about DFT+$U$+$V$ can be found in the pioneering paper~\cite{campo2010extended} as well as in more recent papers~\cite{TancogneDejean:2020, Lee:2020}.
Hence, computing the values of $U^I$ and $V^{IJ}$ interaction parameters is crucial to determine the degree of localization of $3d$ electrons on Mn sites and the degree of hybridization of these $3d$ electrons with $2p$ electrons centered on neighboring O sites. In the next subsection we discuss briefly how these Hubbard parameters can be computed using linear response theory.

\subsection{Calculation of $U$ and $V$}
\label{sec:CalcUV_theory}

First of all, we recall that in Hubbard-corrected DFT the values of Hubbard parameters are not known {\it a~priori}, and hence very often these values are adjusted empirically such that the final results of simulations match some experimental properties of interest. This is fairly arbitrary, as can be seen on the example of $\beta$-MnO$_2$ as was discussed in Sec.~\ref{sec:intro}. Therefore, first-principles calculation of Hubbard parameters for any system at hand is essential and highly desirable. In this paper, we compute $U$ and $V$ from a generalized piece-wise linearity condition imposed through linear-response theory~\cite{cococcioni2005linear}, based on density-functional perturbation theory (DFPT)~\cite{timrov2018hubbard, Timrov:2021}. Within this framework the Hubbard parameters are the elements of an effective interaction matrix computed as the difference between bare and screened inverse susceptibilities~\cite{cococcioni2005linear}:
\begin{equation}
U^I = \left(\chi_0^{-1} - \chi^{-1}\right)_{II} \,,
\label{eq:Ucalc}
\end{equation}
\begin{equation}
V^{IJ} = \left(\chi_0^{-1} - \chi^{-1}\right)_{IJ} \,,
\label{eq:Vcalc}
\end{equation}
where $\chi_0$ and $\chi$ are the susceptibilities which measure the response of atomic occupations to shifts in the potential acting on individual Hubbard manifolds. In particular, $\chi$ is defined as $\chi_{IJ} = \sum_{m\sigma} \left(dn^{I \sigma}_{mm} / d\alpha^J\right)$, where $\alpha^J$ is the strength of the perturbation of electronic occupations of the $J$th site. While $\chi$ is evaluated at self-consistency of the DFPT calculation, $\chi_0$ (which has a similar definition as $\chi$) is computed before the self-consistent re-adjustment of the Hartree and exchange-correlation potentials~\cite{timrov2018hubbard}. In DFPT, the response of the occupation matrix is computed in a primitive unit cell as:
\begin{equation}
\frac{dn^{I \sigma}_{mm'}}{d\alpha^J} = \frac{1}{N_{\mathbf{q}}}\sum_{\mathbf{q}}^{N_{\mathbf{q}}} e^{i\mathbf{q}\cdot(\mathbf{R}_{l} - \mathbf{R}_{l'})}\Delta_{\mathbf{q}}^{s'} n^{s \sigma}_{mm'} \,,
\label{eq:dnq}
\end{equation}
where $\mathbf{q}$ is the wavevector of the monochromatic perturbation, $N_\mathbf{q}$ is the total number of $\mathbf{q}$'s, $\Delta_{\mathbf{q}}^{s'} n^{s \sigma}_{mm'}$ is the lattice-periodic response of atomic occupations to a $\mathbf{q}$-specific monochromatic perturbation, $I\equiv(l,s)$ and $J\equiv(l',s')$ where $s$ and $s'$ are the atomic indices in unit cells while $l$ and $l'$ are the unit cell indices, $\mathbf{R}_l$ and $\mathbf{R}_{l'}$ are the Bravais lattice vectors. The $\mathbf{q}$ grid is chosen fine enough to make the resulting atomic perturbations effectively decoupled from their periodic replicas. More details about this approach can be found in Refs.~\cite{timrov2018hubbard, Timrov:2021}. We stress that the main advantage of using DFPT is that it does not require the usage of computationally expensive supercells contrary to the original linear-response formulation of Ref.~\cite{cococcioni2005linear}. However, it is crucial to remember that the values of the computed Hubbard parameters are strongly dependent on the type of Hubbard projector functions that are used in the DFT+$U$ and DFT+$U$+$V$ approaches; this aspect is discussed in more detail in the next subsection.  

\subsection{Choice of Hubbard projectors}
\label{sec:Hub_projectors_theory}

The Hubbard manifold $\{ \phi_m^I(\mathbf{r}) \}$ can be constructed using different types of projector functions (see e.g. Refs.~\cite{Tablero:2008, Timrov:2020b}). Here we consider two types of projector functions, namely NAO and OAO. NAO is one of the most simple projector functions for the Hubbard manifold, and it is particularly well suited for systems with mostly ionic bonding (because the overlap between orbitals centered on neighboring atoms is small). These projector functions are simply atomic orbitals $\phi_m^I(\mathbf{r})$ that are provided with pseudopotentials, and they are orthonormal within each atom but not between different atoms. However, whenever covalent interactions become important, this type of projector functions is not the best choice since the overlap between orbitals sitting on different atoms becomes significant. In this case, OAO is a better choice. OAO are obtained by taking atomic orbitals of each atom and then orthonormalizing them to all orbitals of all atoms in the system. In this work, we will use the L\"owdin orthogonalization method~\cite{Lowdin:1950, Mayer:2002}, which gives:
\begin{equation}
    \tilde{\phi}^I_{m}(\mathbf{r}) = \sum_{J m'} \left(\hat{O}^{-\frac{1}{2}}\right)^{JI}_{m' m} \phi^J_{m'}(\mathbf{r}) \,,
    \label{eq:OAO_def}
\end{equation}
where $\hat{O}$ is the orbital overlap matrix which is defined as:
\begin{figure}[t]
  \includegraphics[width=0.98\linewidth]{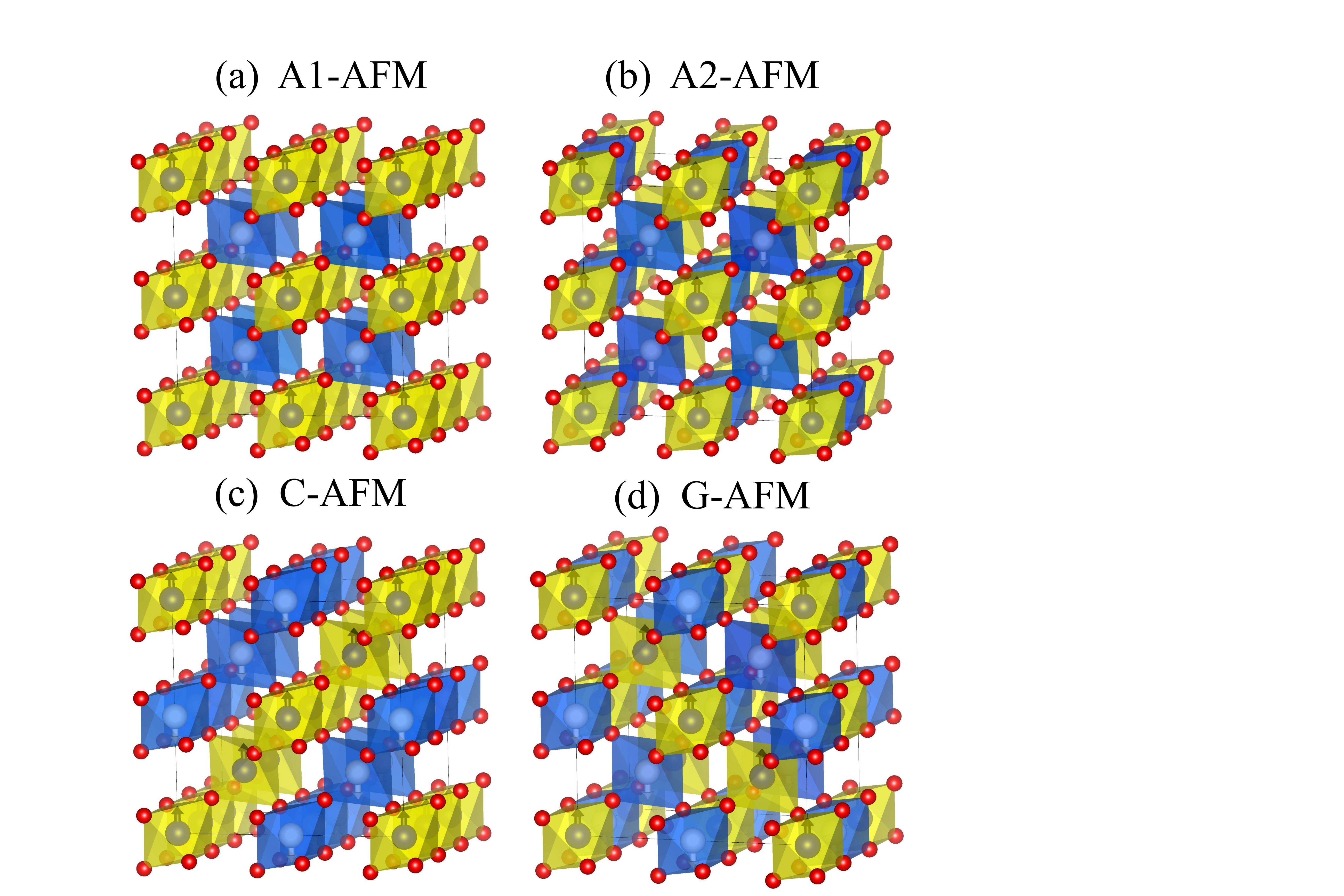}
   \caption{Four collinear AFM configurations of $\beta$-MnO$_2$: (a)~A1-AFM, (b)~A2-AFM, (c)~C-AFM, and (d)~G-AFM. Yellow and blue octahedra contain Mn atoms in the center with up and down spin alignments (shown with black and white arrows), respectively, and oxygen atoms are represented as red balls.}
\label{fig:MagneticStructure}
\end{figure}

\begin{table*}[t]
\renewcommand{\arraystretch}{1.3}
\centering
\begin{tabular}{l|c|c|c|c|c|c|c}
\hline\hline
\multirow{2}{*}{Method}      &  \multirow{2}{*}{\begin{tabular}[c]{@{}c@{}}HP (eV)\end{tabular}}       & \multirow{2}{*}{\begin{tabular}[c]{@{}c@{}}\parbox{1cm}{HM}\end{tabular}}              & \multicolumn{5}{c}{Magnetic ordering}                      \\ \cline{4-8} 
                             &                &                 & \parbox{2cm}{FM}          & \parbox{2cm}{A1-AFM}       & \parbox{2.5cm}{A2-AFM}      & \parbox{3.5cm}{C-AFM}         & \parbox{2.5cm}{G-AFM}     \\ \hline
\multirow{2}{*}{DFT+$U$}     & \multirow{2}{*}{$U$}  & NAO & 4.93          & 4.67           & 4.68               & 4.68, 4.78                 & 4.68               \\ \cline{3-8} 
                             &      & OAO                  & 7.08          & 6.34           & 6.41               & 6.54, 6.64                 & 6.32               \\ \hline
\multirow{4}{*}{DFT+$U$+$V$} & $U$  & \multirow{2}{*}{NAO} & 5.65          & 5.38           & 5.39               & 5.42, 5.62                 & 5.40               \\ \cline{2-2} \cline{4-8} 
                             & $V$  &                      & (1.34, 1.32)  & (1.27, 1.10)   & (1.14, 1.10, 1.23) & (1.13, 1.39), (1.33, 1.31) & (1.16, 1.24, 1.08) \\ \cline{2-8} 
                             & $U$  & \multirow{2}{*}{OAO} & 7.33          & 6.76           & 6.77               & 6.93, 7.03                 & 6.77               \\ \cline{2-2} \cline{4-8} 
                             & $V$  &                      & (1.17, 1.11)  & (0.99, 1.10)   & (0.99, 1.00, 1.07) & (1.05, 1.11), (1.18, 1.07) & (1.01, 0.98, 1.07) \\
\hline\hline
\end{tabular}%
\caption{Hubbard parameters (HP) in eV computed from first principles using DFPT (see Sec.~\ref{sec:CalcUV_theory}), for five magnetic configurations: FM, A1-AFM, A2-AFM, C-AFM, and G-AFM. The onsite $U$ for Mn($3d$) states and intersite $V$ between Mn($3d$) and O($2p$) states are computed in the frameworks of DFT+$U$ and DFT+$U$+$V$ (PBEsol functional) using two types of Hubbard manifolds (HM), NAO and OAO (see Sec.~\ref{sec:Hub_projectors_theory}). In the case of C-AFM there are two values of $U$ because there are two inequivalent types of Mn atoms. Intersite $V$ parameters depend on the distance between Mn and O atoms, therefore there are either two or three values per each case written inside round brackets (depending on how many different bond lengths there are); in the C-AFM case there are two couples of $V$ values refering to two inequivalent Mn atoms.}
\label{tab:Hub_param}
\end{table*}

\begin{equation}
    (\hat{O})^{IJ}_{m_1 m_2} = \braket{\phi^I_{m_1}}{\phi^J_{m_2}} \,,
    \label{eq:overlap_def}
\end{equation}
and $(\hat{O})^{IJ}_{m_1 m_2}$ is a matrix element of $\hat{O}$. By doing so, we obtain Hubbard manifold composed of OAO $\{ \tilde{\phi}_m^I(\mathbf{r}) \}$ that must be used in Eq.~\eqref{eq:occ_matrix_0}. This new basis set better represents hybridizations of orbitals between neighboring sites, but especially it allows us to avoid counting Hubbard corrections twice in the interstitial regions between atoms, which is especially relevant in the case of DFT+$U$+$V$. Finally, it is important to note that NAO and OAO are not truncated at some cutoff radius (at variance with other implementations~\cite{Amadon:2008, Nawa:2018}), which thus eliminates ambiguities due to the choice of such a cutoff radius~\cite{Timrov:2020b, Wang:2016}.

In the following we will see how these two types of Hubbard projectors perform for $\beta$-MnO$_2$ in the framework of DFT+$U$ and DFT+$U$+$V$. 

\section{Technical details}
\label{sec:technical_details}

All calculations were performed using the plane-wave (PW) pseudopotential method as implemented in the \textsc{Quantum ESPRESSO} distribution~\cite{Giannozzi:2009, Giannozzi:2017, Giannozzi:2020}. We have used the xc functional constructed using $\sigma$-GGA with the PBEsol prescription~\cite{Perdew:2008}. Pseudopotentials are taken from the SSSP library~v1.1 (precision)~\cite{prandini2018precision, MaterialsCloud}: For Mn we have used \texttt{mn\_pbesol\_v1.5.uspp.F.UPF} from the GBRV v1.5 library~\cite{Garrity:2014}, and for O we have used \texttt{O.pbesol-n-kjpaw\_psl.0.1.UPF} from the Pslibrary v0.3.1~\cite{Kucukbenli:2014}. We have considered five collinear magnetic orderings: FM, A1-AFM, A2-AFM, C-AFM, and G-AFM; all AFM spin configurations are shown in Fig.~\ref{fig:MagneticStructure}. In the case of FM and A1-AFM the unit cell contains 6 atoms, while in the case of A2-AFM, C-AFM, and G-AFM a supercell of size $2 \times 2 \times 2$ is used containing 48 atoms. For each of these spin configurations, the crystal structure was optimized at three levels of theory (DFT, DFT+$U$, and DFT+$U$+$V$) using the Broyden-Fletcher-Goldfarb-Shanno (BFGS) algorithm~\cite{Fletcher:1987}, with a convergence threshold for the total energy of $10^{-6}$~Ry, for forces of $10^{-5}$~Ry/Bohr, and for pressure of 0.5~Kbar. DFT+$U$ and DFT+$U$+$V$ calculations were performed using two types of Hubbard projectors, NAO and OAO (see Sec.~\ref{sec:methods}). For metallic ground states, we used the Marzari-Vanderbilt (MV) smearing~\cite{Marzari:1999} with a broadening parameter of $5 \times 10^{-3}$~Ry. Structural optimizations were performed using an uniform $\Gamma$-centered $\mathbf{k}$ points mesh of size $8 \times 8 \times 12$ for the 6-atoms unit cell and $4 \times 4 \times 6$ for the 48-atoms supercell, and KS wavefunctions and potentials were expanded in PWs up to a kinetic-energy cutoff of 90 and 1080~Ry, respectively. Total energy differences, magnetic moments, band gaps, and the projected density of states (PDOS) were computed using a more refined $\mathbf{k}$ points mesh of size $12 \times 12 \times 16$ for the 6-atoms unit cell and $6 \times 6 \times 8$ for the 48-atom supercell, and KS wavefunctions and potentials were expanded in PWs up to a kinetic-energy cutoff of 150 and 1800~Ry, respectively. PDOS was plotted using the Gaussian smearing with a broadening parameter of $4.4 \times 10^{-3}$~Ry.

The DFPT calculations of Hubbard parameters were performed using uniform $\Gamma$-centered $\mathbf{k}$ and $\mathbf{q}$ point meshes~\cite{timrov2018hubbard}: for the 6-atoms unit cells (FM and A1-AFM) we used the $4 \times 4 \times 8$ $\mathbf{k}$ points mesh and the $2 \times 2 \times 4$ $\mathbf{q}$ points mesh, while for the 48-atom supercells (A2-AFM, C-AFM, G-AFM) we used the $2 \times 2 \times 4$ $\mathbf{k}$ points mesh and the $1 \times 1 \times 2$ $\mathbf{q}$ points mesh. In DFPT calculations, KS wavefunctions and potentials are expanded in PWs up to a kinetic-energy cutoff of 60 and 720~Ry, respectively. For DFPT calculations on top of metallic ground states, we used the MV smearing with a broadening parameter of $1.5 \times 10^{-2}$~Ry. With this setup, the accuracy of computed Hubbard parameters is 0.01~eV. The DFPT equations are solved using the conjugate-gradient algorithm~\cite{Payne:1992} and the mixing scheme of Ref.~\cite{Johnson:1988}. We have used the self-consistent procedure for the calculation of $U$ and $V$ as described in detail in Ref.~\cite{Timrov:2021}, which consists of cyclic calculations containing structural optimizations and recalculations of Hubbard parameters for each new geometry.

\section{Results and Discussions}
\label{sec:Results_and_Discussion}

\subsection{Hubbard parameters}
\label{sec:Hubbard_parameters}

We start by analyzing the computed Hubbard parameters, which are listed in Table~\ref{tab:Hub_param} for five collinear magnetic configurations. It can be seen that the values of Hubbard parameters $U$ and $V$ are very sensitive to the type of the Hubbard projectors, as is well known from the literature for other materials~\cite{timrov2018hubbard, KirchnerHall:2021, Kick:2019, Nawa:2018}. Due to the use of different Hubbard projectors (NAO or OAO), the variations in the onsite $U$ parameters range from about 1.4 to 2.2~eV, while the variations in the intersite $V$ parameters are not more than 0.3~eV, depending on the magnetic ordering. Therefore, the DFT+$U$ and DFT+$U$+$V$ calculations must be performed consistently, i.e., using the computed Hubbard parameters and the same type of Hubbard projectors that were used to determine these parameters.

Another important observation that can be made by analyzing the data of Table~\ref{tab:Hub_param} is that for a fixed type of Hubbard projectors the variations in the values of Hubbard parameters due to different magnetic orderings is of the order of a fraction of an eV. Similar trends were found in Ref.~\cite{Cococcioni:2021} for molecular systems. This means that the sensitivity of $U$ and $V$ to the spin configuration of $\beta$-MnO$_2$ is not very strong but at the same time it is not negligible (especially in the OAO case).

Using our computed Hubbard parameters for five magnetic orderings of $\beta$-MnO$_2$, we have performed DFT+$U$ and DFT+$U$+$V$ calculations of the structural, electronic, and magnetic properties of this material. Our paper is thus fully first-principles, i.e., without any adjustable or fitting parameters, which allows us to avoid ambiguities in the results that are common to the vast majority of the previous DFT+$U$ studies of $\beta$-MnO$_2$.

\subsection{Structural properties}
\label{sec:Structural_properties}

\begin{figure}[t]
   \includegraphics[width=0.9\linewidth]{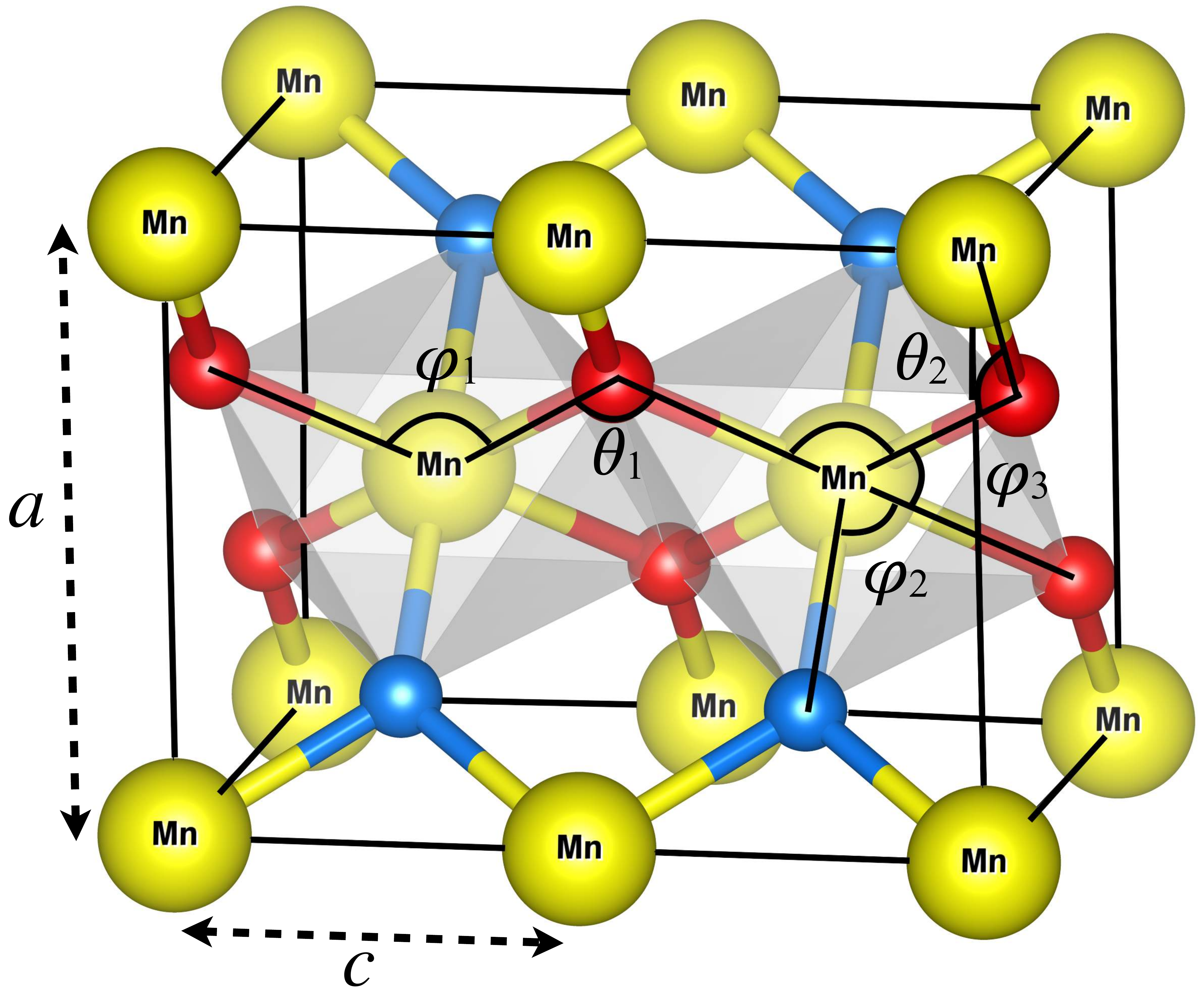}
   \caption{Experimental crystal structure of $\beta$-MnO$_2$~\cite{barudvzija2016structural}. Mn atoms are indicated in yellow color, while O atoms are indicated in blue and red to highlight longer and shorter Mn--O bonds, respectively. Three types of O--Mn--O bond angles are indicated with $\varphi_1$, $\varphi_2$, and $\varphi_3$, while two types of Mn--O--Mn bond angles are indicated with $\theta_1$ and $\theta_2$. $a$ and $c$ are the lattice parameters. The experimental values of bond lengths and bond angles are indicated in Fig.~\ref{fig:Interatomic_dist}.}
\label{fig:crystal_structure}
\end{figure}

$\beta$-MnO$_2$ crystallizes in the tetragonal rutile structure, where the cation (Mn) atoms are octahedrally coordinated to six oxygen atoms (see Fig.~\ref{fig:crystal_structure}). The unit cell with the FM or A1-AFM magnetic orderings consists of two formula units where Mn atoms occupy 2a Wyckoff positions while oxygen atoms occupy 4f Wyckoff positions~\cite{barudvzija2016structural}. 
Instead, A2-AFM, C-AFM, and G-AFM magnetic orderings require $2\times 2 \times 2$ supercells shown in Fig.~\ref{fig:MagneticStructure} (A1-AFM ordering is also shown with a supercell just for the sake of comparison with other AFM orderings). MnO$_6$ octahedra are interconnected primarily along the $c$ axis through edge-sharing, and the remaining octahedra are connected by point sharing with each other. 

First we analyze the optimized lattice parameters $a$ and $c$ and the equilibrium volume $V$ computed at three levels of theory (DFT, DFT+$U$, and DFT+$U$+$V$) with those determined experimentally. As can be seen from Table~\ref{tab:my-table}, standard DFT is in surprisingly good agreement for the lattice parameters and volume when the FM ordering is used ($\Delta V \approx -1.6\%$), while the deviation from the experiment is larger for all types of the AFM ordering ($\Delta V$ ranging from $-2.7$ to $-3.2\%$ depending on the AFM type). Thus, we observe that DFT tends to underestimate lattice parameters and volume, which is in contrast to previous DFT studies~\cite{franchini2007ground}. This difference in the DFT trends can be attributed to the fact that in this paper we use GGA-PBEsol while in Ref.~\cite{franchini2007ground} GGA-PBE was used. We note that despite some uncertainties in the experimental volumes (55.48~\AA$^3$~\cite{Ohama:1971}, 55.82~\AA$^3$~\cite{barudvzija2016structural} ), our conclusions are robust.

\begin{table*}[t]
\centering
\begin{tabular}{c|c|c|c|c|c|c}
\hline\hline
 \multirow{2}{*}{\begin{tabular}[c]{@{}c@{}}\parbox{2cm}{Magnetic \\ ordering} \end{tabular}} & \multirow{2}{*}{\begin{tabular}[c]{@{}c@{}}\parbox{2.8cm}{Crystal structure \\ properties} \end{tabular}}  & \multirow{2}{*}{\parbox{2cm}{DFT}} & \multicolumn{2}{c|}{NAO} & \multicolumn{2}{c}{OAO}\\ \cline{4-7} 
   & &  & \parbox{2cm}{DFT+$U$} & \parbox{2cm}{DFT+$U$+$V$} & \parbox{2cm}{DFT+$U$} & \parbox{2cm}{DFT+$U$+$V$}   \\ \hline
\multirow{4}{*}{FM} & $a$ (\AA) & 4.40	& 4.42	& 4.39 & 4.49 & 4.43  \\ 
 & $c$ (\AA) & 2.84 & 2.97 & 	2.89& 3.05 & 2.97 \\ 
  & $V$ (\AA$^3$) &  54.94	& 58.23	& 55.82	& 61.41	& 58.32  \\
 & $\Delta V$ $\%$  &   -1.58 & 	4.32 & 	0	&  10.01 & 	4.48   \\
 \hline
 
\multirow{4}{*}{A1-AFM} & $a$ (\AA) & 4.37  &  4.38	& 4.36	& 4.40 &	4.39 \\ 
 & $c$ (\AA) &2.83	& 2.93 & 2.89 &	2.94 & 2.92 \\ 
  & $V$ (\AA$^3$) &  54.06	&  56.23	&  55.03 & 	57.07	&  56.35    \\
 & $\Delta V$ $\%$ &  -3.15	& 0.73	& -1.42	& 2.24 &	0.95  \\
 \hline
 
\multirow{4}{*}{A2-AFM} & $a$ (\AA) & 4.38	& 4.39	& 4.37 & 4.41 &	4.39  \\ 
 & $c$ (\AA) & 2.83 & 2.93 & 2.89 & 2.95 &  2.92  \\ 
 & $V$ (\AA$^3$) &  54.23	 & 56.30	  & 55.13  &	57.50	 & 56.43    \\
 & $\Delta V$ $\%$ &   -2.85 &	0.86 &	-1.24	& 3.01 &	1.09   \\
 \hline

\multirow{4}{*}{C-AFM} & $a$ (\AA) & 4.38	& 4.40	& 4.37 &	4.44 &	4.41	 \\  
 & $c$ (\AA) & 2.83 & 	2.94	&  2.89 & 	2.98	& 2.93\\
   & $V$ (\AA$^3$) &  54.34	& 56.96 &	55.27	& 58.75	& 57.07  \\
& $\Delta V$ $\%$  & -2.65	& 2.04	& -0.99	& 5.25	& 2.24  \\
 \hline

\multirow{4}{*}{G-AFM} & $a$ (\AA) & 4.38	& 4.39 &	4.37	& 4.40 &	4.39  \\  
 & $c$ (\AA) & 2.83	& 2.93 &	2.89	& 2.95	& 2.92 \\ 
   & $V$ (\AA$^3$) & 54.22	& 56.31 &	55.14	& 57.19	& 56.44   \\
& $\Delta V$ $\%$ &   -2.87	& 0.88	& -1.22	& 2.45 & 1.11 \\

 \hline\hline
\end{tabular}
\caption{Crystal structure properties of $\beta$-MnO$_2$ (see Fig.~\ref{fig:crystal_structure}): lattice parameters $a$ and $c$ (in \AA), volume $V$ (in \AA$^3$) corresponding to the 6-atoms unit cell and its deviation $\Delta V$ (in $\%$) from the experimental one of Ref.~\cite{barudvzija2016structural}. The results are presented for five collinear magnetic orderings (FM, A1-AFM, A2-AFM, C-AFM, G-AFM) computed at three level of theory: DFT, DFT+$U$, and DFT+$U$+$V$ (PBEsol functional). Hubbard-corrected results are presented when using two types of Hubbard projectors (NAO and OAO) with their respective Hubbard parameters (see Table~\ref{tab:Hub_param}). Experimental lattice parameters for the noncollinear helical magnetic ordering of $\beta$-MnO$_2$ are $a_\mathrm{exp}=4.41$~\AA \, and $c_\mathrm{exp}=2.87$~\AA, and the experimental volume is $V_\mathrm{exp}=55.82$~\AA$^3$ according to Ref.~\cite{barudvzija2016structural}; $a_\mathrm{exp}=4.40$~\AA, $c_\mathrm{exp}=2.88$~\AA \, according to Ref.~\cite{bolzan1993powder}; $V_\mathrm{exp}=55.48$~\AA$^3$ according to Ref.~\cite{Ohama:1971}.}
\label{tab:my-table}
\end{table*}
Hubbard-corrected DFT results vary significantly depending on whether only $U$ or both $U$ and $V$ are included, on the type of Hubbard projectors, and on the magnetic ordering. In particular, as can be seen in Table~\ref{tab:my-table} we find that using NAO projectors DFT+$U$ generally tends to overestimate volumes while DFT+$U$+$V$ tends to underestimate them; lattice parameter $a$ is generally in good agreement with the experimental one while $c$ is overestimated especially at the DFT+$U$ level. Instead, using OAO we find that both DFT+$U$ and DFT+$U$+$V$ overestimate the volume and the lattice parameter $c$ (especially at the DFT+$U$ level), while $a$ is either slightly overestimated or agrees remarkably with the experimental value. Overall, we can conclude that DFT+$U$+$V$ is in better agreement with the experiments than DFT+$U$ when OAO projectors are used; in terms of magnetic ordering, A1-AFM comes out to be the most accurate representation. In the case of NAO, the best agreement with experiments is obtained at the DFT+$U$+$V$ level for the FM ordering. This latter finding has a drawback that FM is not expected to be a good approximation of the noncollinear helical magnetic ordering. 
Therefore, next best results obtained using NAO are for A1-AFM, A2-AFM, and G-AFM using DFT+$U$, all being very similar on average in terms of accuracy. Hence, we find that there is no one single case which agrees best with experiments simultaneously for the lattice parameters and volume, but the most accurate results are obtained for A1-AFM at DFT+$U$+$V$ (OAO), A1-AFM and A2-AFM at DFT+$U$ (NAO).

\begin{figure}[t]
  \includegraphics[width=0.95\linewidth]{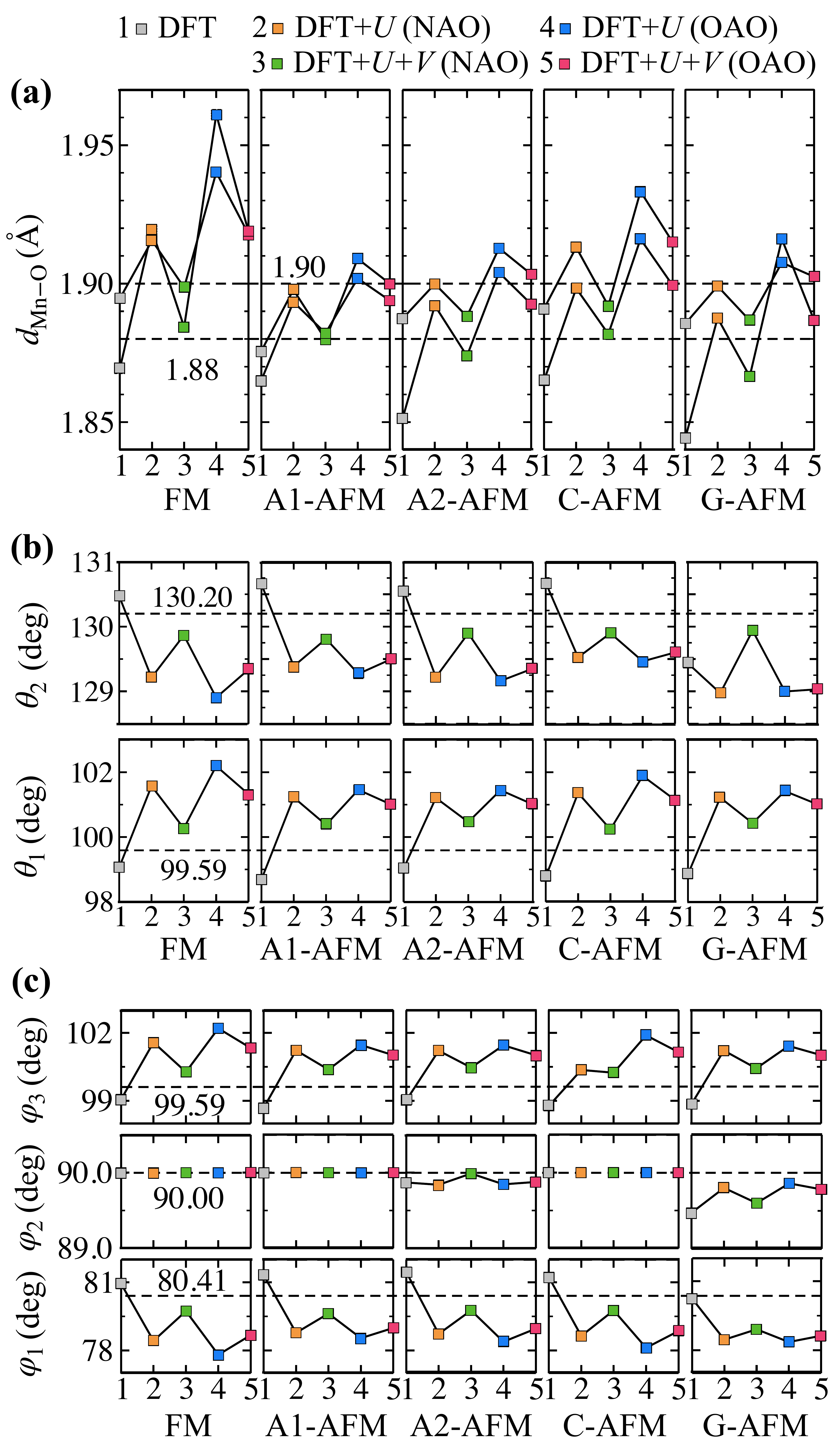}
\caption{Bond lengths and bond angles in $\beta$-MnO$_2$ (see Fig.~\ref{fig:crystal_structure}) as computed in this work using DFT, DFT+$U$, and DFT+$U$+$V$ (PBEsol functional) with two types of Hubbard projectors (NAO and OAO) and corresponding first-principles parameters $U$ and $V$ (see Table~\ref{tab:Hub_param}), and as measured in experiments (dashed horizontal lines)~\cite{barudvzija2016structural}. Theoretical results are shown for five collinear magnetic orderings (FM, A1-AFM, A2-AFM, C-AFM, G-AFM), while the experimental results correspond to the noncollinear helical structure. (a)~Mn-O bond lengths (in \AA), (b)~Mn-O-Mn bond angles $\theta_1$ and $\theta_2$ (in degrees), and (c)~O-Mn-O bond angles $\varphi_1$, $\varphi_2$, and $\varphi_3$ (in degrees). Solid lines that connect squares are guides to the eye.}
\label{fig:Interatomic_dist}
\end{figure}

Now we turn to the analysis of the Mn-O bond lengths ($d_\mathrm{Mn-O}$) and two types of bond angles, namely Mn-O-Mn (labeled as $\theta_1$ and $\theta_2$) and O-Mn-O (labeled as $\varphi_1$, $\varphi_2$, and $\varphi_3$), as shown in Fig.~\ref{fig:crystal_structure}. The optimized and experimental values of these quantities are highlighted in Fig.~\ref{fig:Interatomic_dist}. Experimentally it was observed that there are two types of Mn-O interatomic distances (see Fig.~\ref{fig:crystal_structure}): two shorter distances of length 1.88~\AA \, and one longer distance of length 1.90~\AA~\cite{barudvzija2016structural}. Since the difference between these two values is very small, it is not easy to resolve them accurately in calculations as is shown in the following. As seen in Fig.~\ref{fig:Interatomic_dist}~(a), for the FM ordering the closest agreement with these two experimental distances is obtained using DFT+$U$+$V$ (NAO), for A1-AFM using DFT+$U$+$V$ (OAO), for A2-AFM and G-AFM  using DFT+$U$ (NAO), and for C-AFM using DFT+$U$+$V$ (NAO). Thus, depending on the type of the magnetic ordering we find that different levels of theory perform better or worse than the others. Importantly, we find that in the majority of cases DFT gives the worst result for $d_\mathrm{Mn-O}$. As for what concerns Mn-O-Mn bond angles, the experimental values are $\theta_1 = 99.59^\circ$ and $\theta_2 = 130.20^\circ$~\cite{barudvzija2016structural}. It can be seen from Fig.~\ref{fig:Interatomic_dist}~(b) that the closest agreement with these values is obtained either using standard DFT or DFT+$U$+$V$ (NAO), depending on the magnetic ordering; other levels of theory give angles $\theta_1$ that $\theta_2$ that are somewhat more overestimated and underestimated, respectively. Interestingly, the trends in the performance of different levels of theory are essentially the same for all considered magnetic orderings. Finally, the O-Mn-O bond angles experimentally have the following values: $\varphi_1 = 80.41^\circ$, $\varphi_2 = 90.00^\circ$, and $\varphi_3 = 99.59^\circ$~\cite{barudvzija2016structural}. In Fig.~\ref{fig:Interatomic_dist}~(c) we see that the closest agreement with the experimental values of $\varphi_1$ and $\varphi_3$ is obtained again either using standard DFT or DFT+$U$+$V$ (NAO). Instead, for $\varphi_2$ we find that excellent agreement with the experimental value is obtained at all levels of theory for FM, A1-AFM, and C-AFM; while for A2-AFM the best agreement is obtained using DFT+$U$+$V$ (NAO), and for G-AFM using DFT+$U$ (OAO). We note however, that the variations in $\varphi_2$ for A2-AFM are extremely small, while for G-AFM they are relatively large (but not more than $0.5^\circ$). Therefore, overall we can conclude that the best agreement with the experimental bond lengths and angles is found for the FM ordering at the DFT+$U$+$V$ (NAO) level of theory. However, as was pointed out above, FM is not expected to be a good approximation to the real noncollinear helical magnetic ordering. Fortunately, the agreement with experiments for some AFM orderings is also very satisfactory at different levels of theory. 

The overall analysis of the structural properties of $\beta$-MnO$_2$ has revealed that the FM ordering turns out to be in best agreement with experiments using DFT+$U$+$V$ (NAO). This is, although, quite surprising since the FM ordering is not the true magnetic ground state of $\beta$-MnO$_2$ and it is not expected to be a good approximation to the true noncollinear helical ordering. Moreover, we found that overall satisfactory results for the structural parameters are also found, in particular, for A1-AFM using DFT+$U$+$V$ (OAO). The accuracy of DFT+$U$+$V$ for predicting structural properties will likely increase further if considering the noncollinear helical ordering. In order to further compare the accuracy of the considered levels of theory for $\beta$-MnO$_2$, we proceed to investigate the relative stability of the five magnetic configurations.

\subsection{Energetics}
\label{sec:Energetics}

\begin{figure}[t]
  \includegraphics[width=0.98\linewidth]{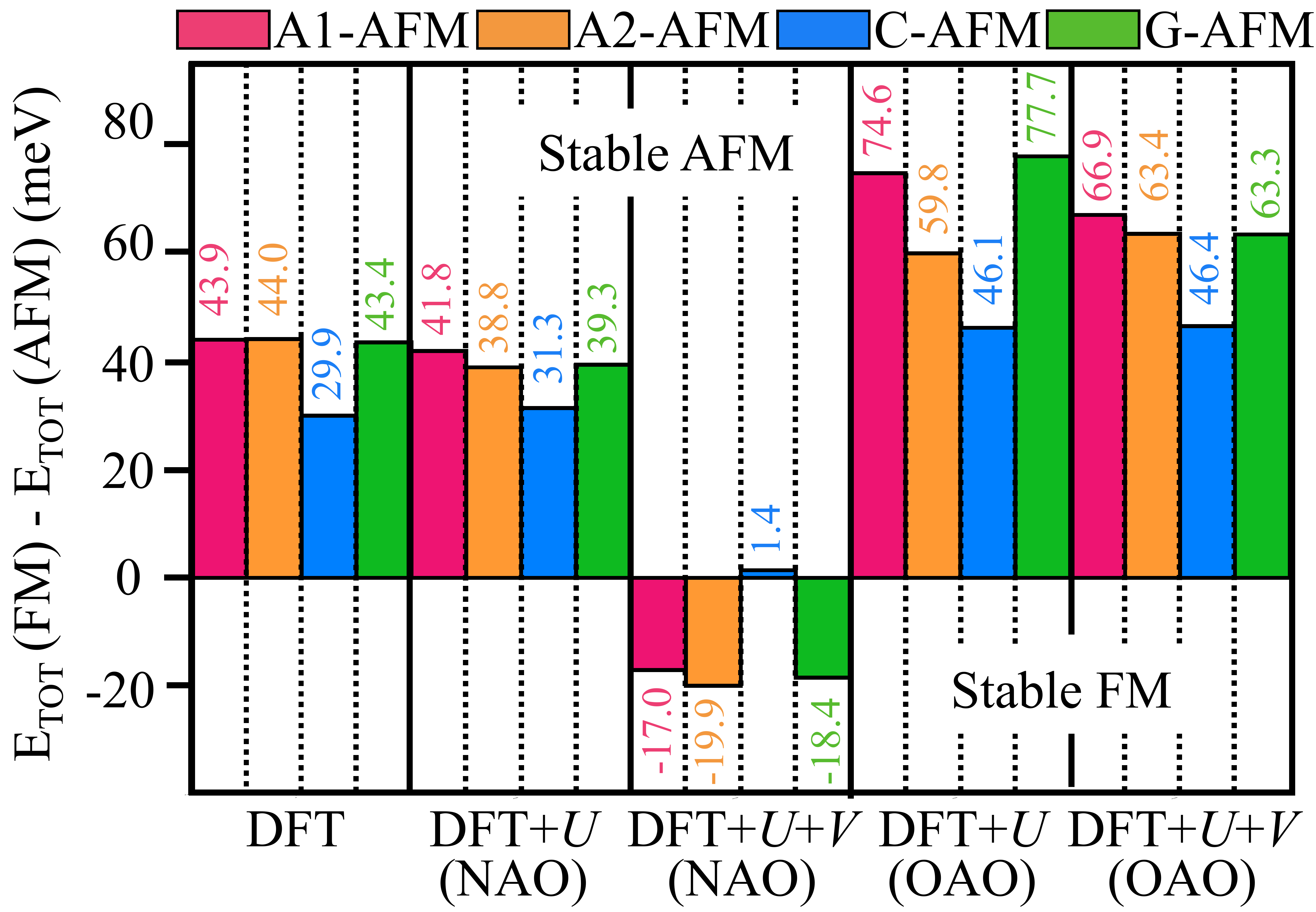} 
  \caption{Total energy difference per formula unit (in meV) for five collinear magnetic orderings (FM, A1-AFM, A2-AFM, C-AFM, G-AFM) computed at three levels of theory (DFT, DFT+$U$, and DFT+$U$+$V$) using the PBEsol functional. For each case, the Hubbard parameters $U$ and $V$ were computed using DFPT, using two types of Hubbard projectors (NAO and OAO), and they are listed in Table~\ref{tab:Hub_param}.}
\label{fig:RE}
\end{figure}

In this section we compare the total energies per formula unit of $\beta$-MnO$_2$ for five collinear magnetic orderings. This comparison is done using standard DFT and Hubbard-corrected DFT, and the result is shown in Fig.~\ref{fig:RE}. First of all, it is important to discuss whether it is correct to compare the total energies for different magnetic orderings computed using different values of the Hubbard parameters. Clearly, when the Hubbard parameters are chosen empirically and arbitrarily, it is not appropriate to compare total energies computed using these parameters~\cite{franchini2007ground}. For this reason, it is common in literature to fix $U$ and compare the total energies. However, when Hubbard parameters are computed from first principles for a given material in different magnetic configurations, it is physically sound to compare the total energies with respective values of $U$ and $V$. Indeed, Hubbard parameters are \textit{not} universal parameters that can be applied globally to all magnetic configurations of a given material or to different compounds containing the same transition-metal element. Hubbard parameters are material-specific. In this paper, $U$ and $V$ are defined as the response property of a material, namely they are computed by perturbing slightly the electronic occupations on Hubbard atoms and by subsequently measuring the (linear) response of the system to such a perturbation~\cite{cococcioni2005linear, timrov2018hubbard, Timrov:2021}. Obviously, this response is not expected to be exactly the same in different compounds or for different spin configurations; as a result, Hubbard parameters change depending on many factors as can be seen in Table~\ref{tab:Hub_param}. In fact, it has been shown that the \textit{ab~initio} value of $U$ can change by a fraction of eV or several eV for the same element in different chemical environments (i.e., depending on the oxidation state)~\cite{Floris:2020, Bennett:2019}. Comparison of energies computed using different Hubbard parameters (computed from first principles self-consistently for each material of interest) was done e.g. in Refs.~\cite{Ricca:2019, Ricca:2020, Cococcioni:2019, Cococcioni:2021}. Therefore, we adopt the same strategy and apply it in this paper for $\beta$-MnO$_2$. In addition it is useful to mention about other recent ideas of how to compare total energies of different spin configurations (namely, comparing plain DFT total energies computed on top of the DFT+$U$ or DFT+$U$+$V$ charge densities~\cite{Mariano:2021}), but this path will not be explored in this paper.    
 
As can be seen from Fig.~\ref{fig:RE}, the prediction of the most energetically stable magnetic configuration depends strongly on the level of theory that is used. We find that at the DFT level all AFM orderings are more energetically favorable than the FM ordering. Among them, A1-AFM and A2-AFM are the lowest in energy and are essentially degenerate, while G-AFM is only slightly higher in energy. We note that the energy differences are extremely small, and this is why it is crucial to converge the calculations with very high accuracy and, in particular, keep high precision in the values of the computed Hubbard parameters. In DFT+$U$ with NAO and OAO projectors, all AFM configurations are more energetically stable than FM, similarly to DFT. A1-AFM turns out to be the most favorable at the DFT+$U$ (NAO) level, while G-AFM is the most stable when DFT+$U$ (OAO) is used. We find that DFT+$U$+$V$ results depend very strongly on the type of projectors that are used; DFT+$U$+$V$ (NAO) predicts that FM is more stable than A1-AFM, A2-AFM, and G-AFM, and it is essentially degenerate in energy with C-AFM (which turns out to be only marginally more stable than FM). Instead, in DFT+$U$+$V$ (OAO) all AFM orderings are more stable than FM, with A1-AFM being the most energetically favorable magnetic ordering. 

Therefore, the commonly accepted A1-AFM spin configuration for $\beta$-MnO$_2$ is found to be the most energetically favorable only at the DFT+$U$ (NAO) and DFT+$U$+$V$ (OAO) levels of theory. It is also important to note that even though DFT+$U$+$V$ (NAO) predicted very accurately the lattice parameters, bond lengths, and bond angles, it completely fails in stabilizing the A1-AFM spin configuration and hence DFT+$U$+$V$ (NAO) cannot be considered as the most accurate computational approach for describing $\beta$-MnO$_2$. This finding suggests that the inclusion of $V$ is not enough and a careful choice of Hubbard projects is required: NAO projectors do not provide the correct energetics of $\beta$-MnO$_2$ within DFT+$U$+$V$, and hence the orthogonalization of atomic orbitals is crucial.

\subsection{Magnetic moment}

\begin{figure}[t]
  \includegraphics[width=0.98\linewidth]{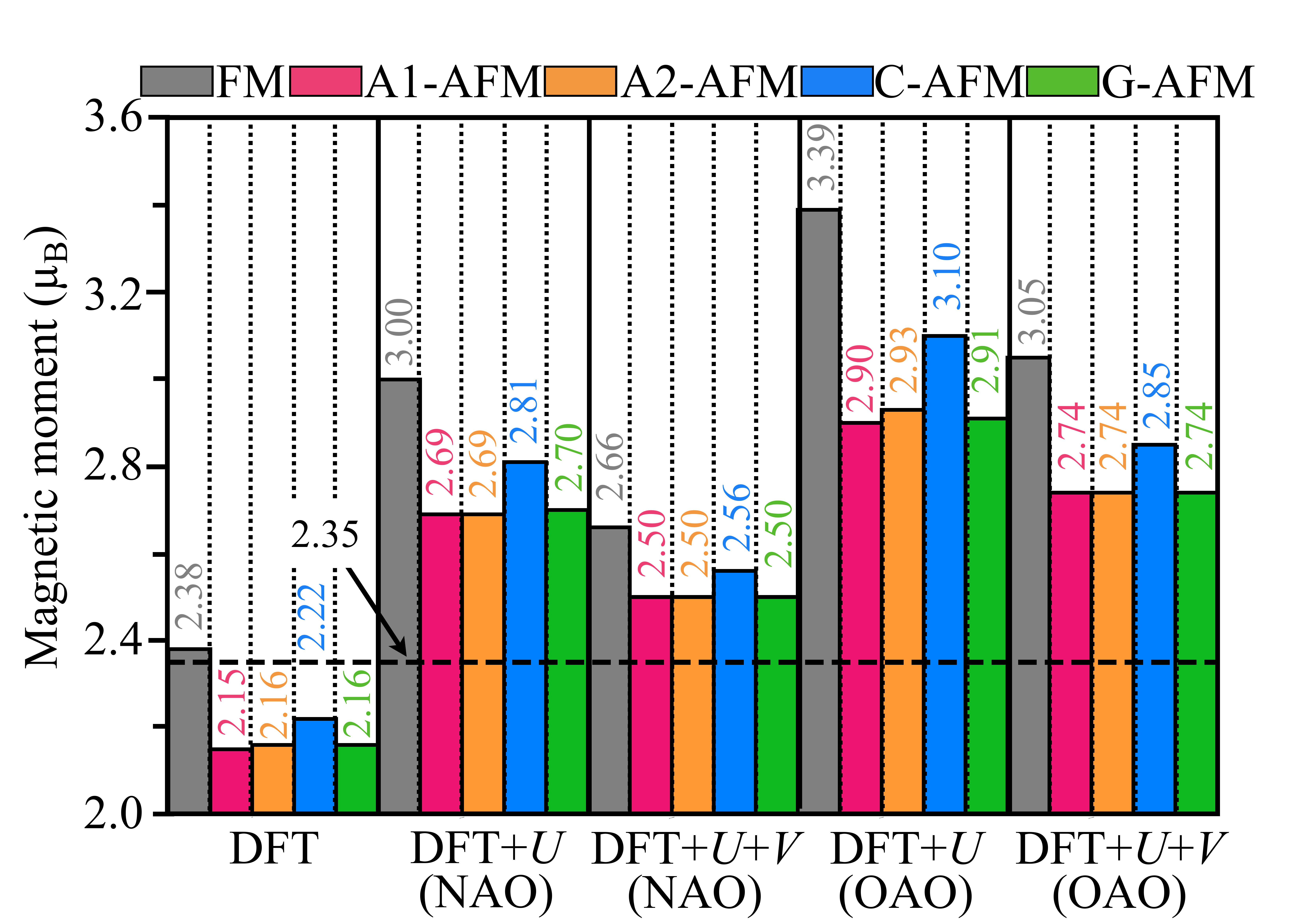} 
\caption{Magnetic moment (in $\mu_\mathrm{B}$) on Mn atoms in $\beta$-MnO$_2$ for five collinear magnetic orderings (FM, A1-AFM, A2-AFM, C-AFM, G-AFM) computed at three levels of theory (DFT, DFT+$U$, and DFT+$U$+$V$) using the PBEsol functional. For each case, the Hubbard parameters $U$ and $V$ were computed using DFPT, using two types of Hubbard projectors (NAO and OAO), and they are listed in Table~\ref{tab:Hub_param}. The experimental magnetic moment at $T=10$~K is 2.35~$\mu_\mathrm{B}$~\cite{regulski2003incommensurate} and it is indicated with a horizontal dashed line.}
\label{fig:Magnetic_moment}
\end{figure}

The magnetic moments on Mn atoms in $\beta$-MnO$_2$ for all five magnetic orderings are shown in Fig.~\ref{fig:Magnetic_moment}. The experimental value of the magnetic moment at $T=10$~K is 2.35~$\mu_\mathrm{B}$~\cite{regulski2003incommensurate}. At the DFT level the magnetic moment is underestimated by about $0.1-0.2$~$\mu_\mathrm{B}$ for all AFM orderings, while the magnetic moment of FM is in very good agreement with the experimental value (with only a slight overestimation of 0.03~$\mu_\mathrm{B}$). Application of the Hubbard corrections leads to the overestimation of the magnetic moment for all cases shown in Fig.~\ref{fig:Magnetic_moment}. In particular, the closest agreement with the experimental value on average is obtained when using DFT+$U$+$V$ (NAO), while the largest deviation is obtained on average when using DFT+$U$ (OAO). The A1-AFM spin configuration that was found to be the most energetically favorable at the DFT+$U$ (NAO) and DFT+$U$+$V$ (OAO) levels of theory (see Sec.~\ref{sec:Energetics}) gives the magnetic moments of 2.69~$\mu_\mathrm{B}$ and 2.74~$\mu_\mathrm{B}$, respectively, that are in fair agreement with the experimental value. It is worth noting that the overestimation of the magnetic moment for A1-AFM was also obtained in previous works using the HSE06 and PBE0 hybrid functionals (giving the magnetic moment of 2.89~$\mu_\mathrm{B}$)~\cite{franchini2007ground} and SCAN meta-GGA functional (2.62~$\mu_\mathrm{B}$)~\cite{gautam2018evaluating}. The overestimation of the magnetic moment in this paper using Hubbard-corrected DFT and in previous papers using hybrid functionals and SCAN might, in part, be attributed to the fact that these calculations were performed for the collinear A1-AFM magnetic ordering which is just an approximation to the real noncollinear helical ordering; thus, perfect agreement with the experimental magnetic moment should not be expected. In addition, the inclusion of the Hund's $J$ in the Hubbard-corrected DFT might further improve the magnetic moments~\cite{Himmetoglu:2011, Bajaj:2017, Linscott:2018}. 

\subsection{Band gap}
\label{sec:Band_gap}

$\beta$-MnO$_2$ is a small band gap semiconductor with the experimental value ranging from 0.26 to 0.28~eV~\cite{Chevillot:1959, yu2008electronic, Islam:2005, druilhe1967electron}. Previous works based on hybrid functionals for A1-AFM predicted a gap of 0.6~eV~\cite{franchini2007ground} and 1.75~\cite{kitchaev2016energetics} (using HSE06), or 1.5~eV~\cite{franchini2007ground} (using PBE0), which largely overestimate the experimental value. Earlier DFT+$U$ works often relied on the empirical $U$ (and $J$) values aiming at reproducing well the experimental band gap. For example, in Ref.~\cite{lim2016improved} the authors used empirical $U=2.8$~eV and $J=1.2$~eV with NAO projectors for the helical magnetic ordering giving the band gap of 0.25~eV, while in Ref.~\cite{tompsett2012importance} the authors used first-principles $U=6.7$~eV and $J=1.2$~eV in the LAPW framework for the A1-AFM ordering giving 0.8~eV (while for the helical ordering the gap differs only by 0.1~eV with respect to the A1-AFM case~\cite{tompsett2012importance}). In other DFT+$U$ works (that do not include $J$ explicitly) an empirical $U$ was used with PAW Hubbard projectors; with $U=4$~eV either a half-metallic FM solution was obtained~\cite{franchini2007ground} (which was due to a too large Gaussian smearing of 0.60~eV~\cite{noda2016momentum}), while in other work~\cite{noda2016momentum} with the same $U$ a semiconducting A1-AFM solution was obtained (using a smaller Gaussian smearing of 0.01~eV) with a largely underestimated band gap of 0.04~eV. On the other hand, meta-GGA studies of A1-AFM using the SCAN functional reported a gap of 0.43~eV~\cite{kitchaev2016energetics}, while a more resent study using SCAN~\cite{gautam2018evaluating} reported that the system is metallic provided that a more dense $\mathbf{k}$ points mesh is used than in Ref.~\cite{kitchaev2016energetics}, while SCAN+$U$ with the empirical $U=2.7$~eV and using PAW projectors gives a gap of 0.64~eV~\cite{gautam2018evaluating}. However, the band gaps in other types of the collinear magnetic ordering of $\beta$-MnO$_2$ were never investigated, and moreover the effect of intersite $V$ interactions on the band gap remained unexplored until now.

Key question is: Can Hubbard-corrected DFT (DFT+$U$ and DFT+$U$+$V$) accurately predict band gaps even if DFT is not a theory for spectral properties? As was shown in Ref.~\cite{KirchnerHall:2021}, DFT+$U$ with $U$ computed using linear-response theory often significantly improves the agreement with the experimental band gaps (in contrast to standard DFT) provided that the Hubbard correction acts on the edge states (if the system is already insulating at the DFT level) or states around the Fermi level (if the system is unphysically metallic at the DFT level) providing a Koopmans-like linearization~\cite{Nguyen:2018}. Moreover, it was found that the values of the band gaps are very sensitive to the type of Hubbard projectors that are used (which influence also the values of the corresponding first-principles Hubbard parameters), with DFT+$U$ (OAO) giving more accurate band gaps than DFT+$U$ (NAO). These observations were very useful to conduct e.g. a DFT+$U$-based high-throughput search of novel materials for the photocatalytic water splitting~\cite{Xiong:2021}. In this paper, we go further and explore the accuracy of the extended DFT+$U$+$V$ formulation for predicting band gaps and its sensitivity to the type of Hubbard projectors. The detailed discussion about the PDOS in $\beta$-MnO$_2$ will be given in Sec.~\ref{sec:PDOS}; here we just remark that the Hubbard $U$ correction is applied only to Mn($3d$) states while no $U$ correction was applied to O($2p$) states [and intersite $V$ correction was applied between Mn($3d$) and O($2p$) in the DFT+$U$+$V$ framework] (see Sec.~\ref{sec:Hubbard_parameters}). 

\begin{figure}[t]
  \includegraphics[width=0.98\linewidth]{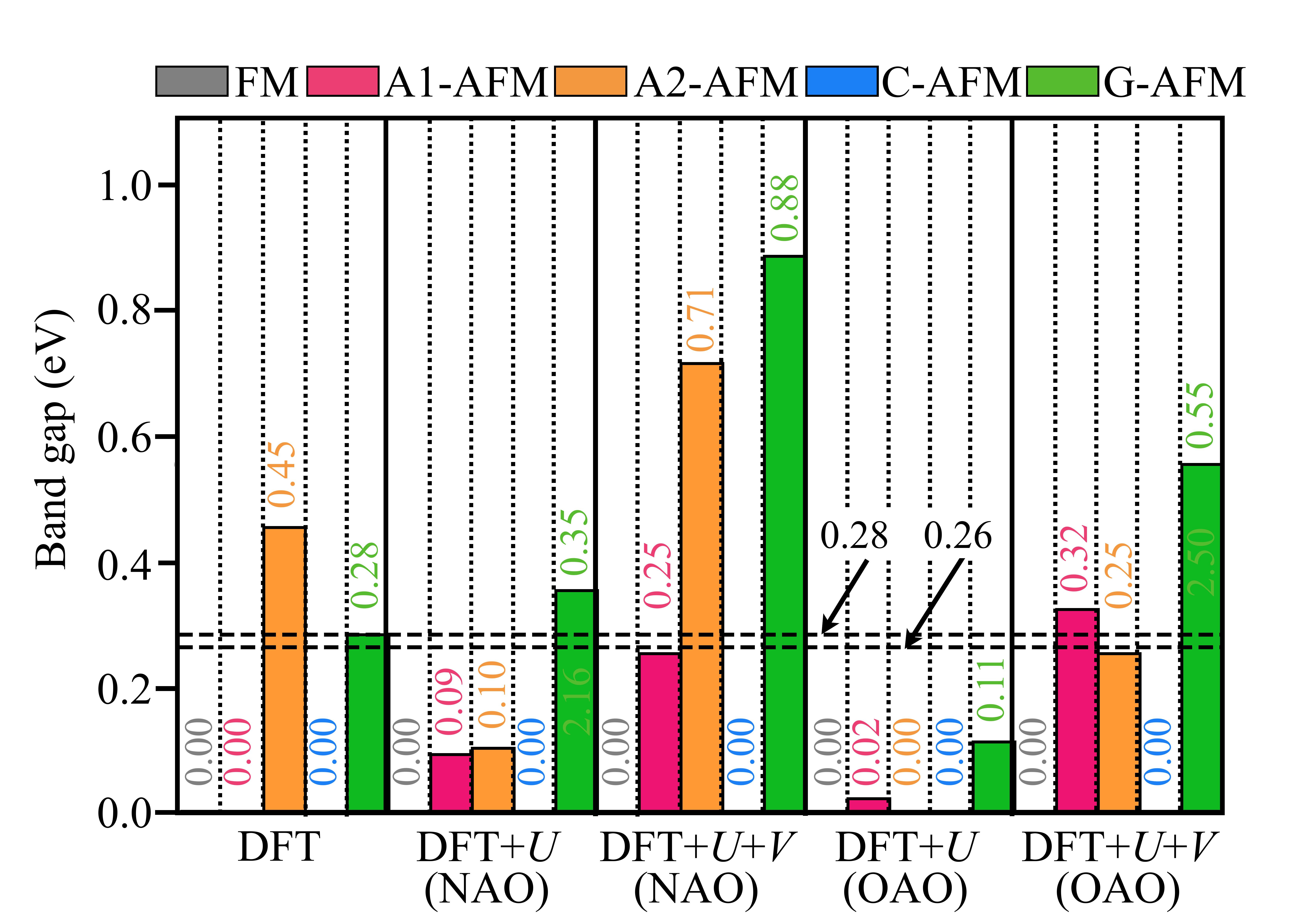}
  \caption{Band gap (in eV) in $\beta$-MnO$_2$ for five collinear magnetic orderings (FM, A1-AFM, A2-AFM, C-AFM, G-AFM) computed at three levels of theory (DFT, DFT+$U$, and DFT+$U$+$V$) using the PBEsol functional. For each case, the Hubbard parameters $U$ and $V$ were computed using DFPT, using two types of Hubbard projectors (NAO and OAO), and they are listed in Table~\ref{tab:Hub_param}. The experimental band gap is 0.26~eV~\cite{Chevillot:1959, yu2008electronic}, 0.27~eV~\cite{Islam:2005}, 0.28~eV~\cite{druilhe1967electron}; the minimum and maximum values are indicated by the horizontal dashed lines.}
\label{fig:Band_gaps}
\end{figure}

The computed band gaps in $\beta$-MnO$_2$ for five collinear magnetic configurations and using DFT with and without Hubbard corrections are shown in Fig.~\ref{fig:Band_gaps}. Overall we can see that in many cases the band gap is zero meaning that some methods predict a metallic unphysical ground state. In particular, at the DFT level of theory only A2-AFM and G-AFM magnetic orderings favor a semiconducting solution,  with G-AFM giving a gap of 0.28~eV which is surprisingly in excellent agreement with the experimental gap. In DFT+$U$ (NAO), A1-AFM and A2-AFM both underestimate the band gap and give very similar values of about $0.1$~eV, while G-AFM overestimates the gap by $0.07-0.09$~eV. In contrast, in DFT+$U$ (OAO), A2-AFM gives a metallic solution while A1-AFM predicts a tiny gap of 0.02~eV, and G-AFM also underestimates the gap significantly (by $0.15-0.17$~eV). Therefore, at the DFT+$U$ level we can see that OAO performs worse than NAO, in variance with what was observed for nonmagnetic materials in Ref.~\cite{KirchnerHall:2021}. This finding suggests that most likely the hybridization effects in $\beta$-MnO$_2$ are important and hence intersite $V$ interactions should be taken into account. Indeed, as can be seen in Fig.~\ref{fig:Band_gaps}, DFT+$U$+$V$ gives much more accurate band gaps for A1-AFM with both NAO and OAO projectors. However, A2-AFM gives an accurate prediction of the band gap only at the DFT+$U$+$V$ (OAO) level, while at the DFT+$U$+$V$ (NAO) level the gap is largely overestimated. This means that not only the intersite $V$ is important, but also the type of projectors must be carefully chosen, with OAO being more accurate than NAO within DFT+$U$+$V$. Regarding G-AFM, the gap is overestimated at DFT+$U$+$V$ for both types of Hubbard projectors.

Therefore, we find that even though DFT predicts surprisingly accurate structural properties and magnetic moment for the FM ordering, it completely fails to predict a semiconducting ground state and to stabilize FM with respect to AFM orderings. On the other hand, DFT+$U$+$V$ (NAO) describes accurately the structural properties, magnetic moment, and band gap for A1-AFM ordering; however, this spin configuration is less energetically favorable than FM at this level of theory. Conversely, DFT+$U$ (NAO) and DFT+$U$+$V$ (OAO) predict A1-AFM to be the most energetically stable and they describe fairly accurately all properties for this magnetic ordering. However, DFT+$U$+$V$ (OAO) predicts the band gap more accurately than DFT+$U$ (NAO) for A1-AFM, thus overall DFT+$U$+$V$ (OAO) comes out to be the most accurate computational framework for $\beta$-MnO$_2$. Finally, the DFT+$U$+$V$ (OAO) band gap is 0.32~eV which is a result that is superior to those obtained using hybrid functionals that give band gaps by a factor $2-6$ worse~\cite{franchini2007ground, kitchaev2016energetics} than the DFT+$U$+$V$ (OAO) band gap when compared to the experimental one. The accuracy of the DFT+$U$+$V$ (OAO) bad gap will likely increase further when considering the helical magnetic ordering of $\beta$-MnO$_2$.

\subsection{Projected density of states}
\label{sec:PDOS}

\begin{figure}[t]
   \includegraphics[width=0.99\linewidth]{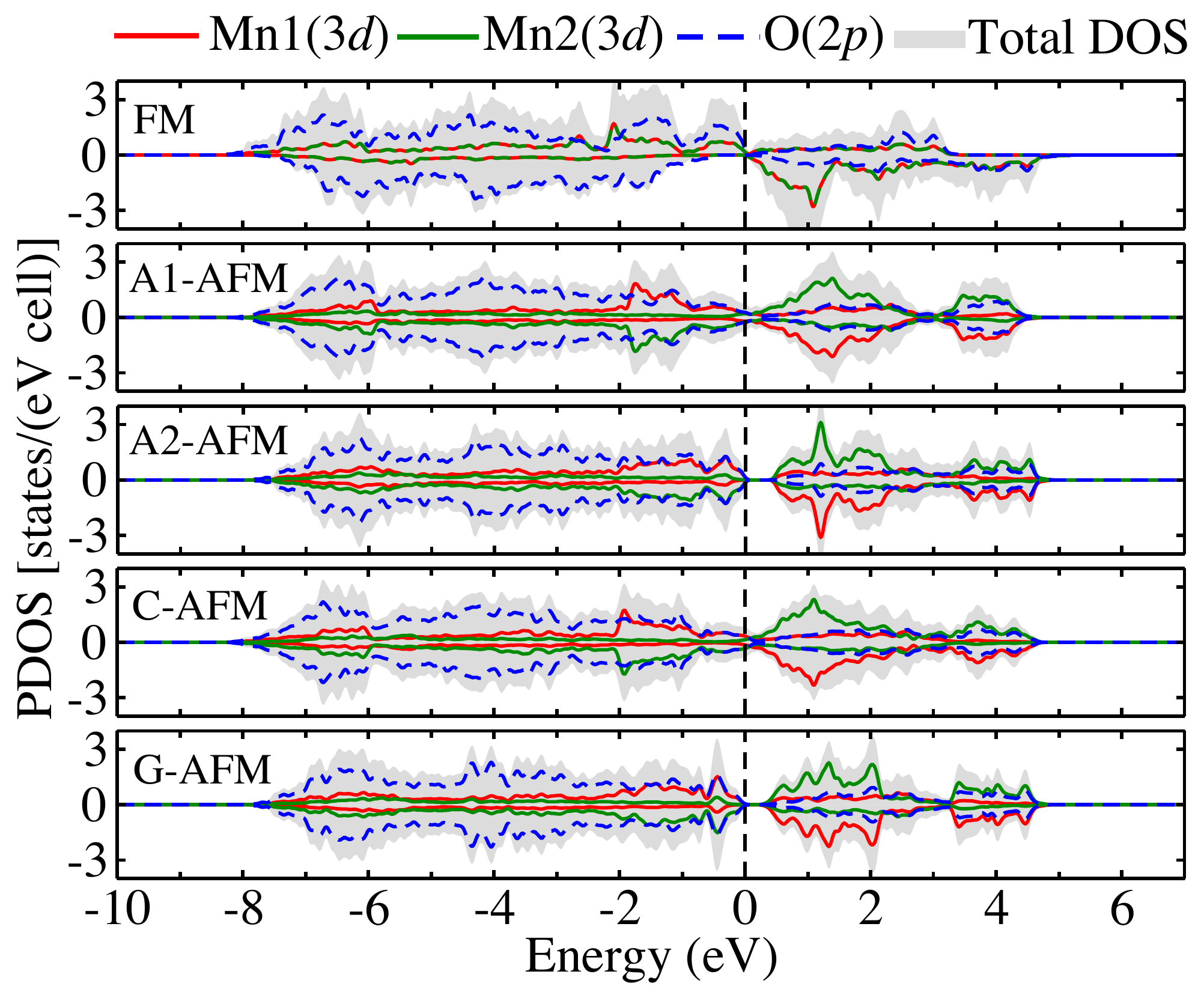}
   \caption{Spin-resolved PDOS and total DOS for five collinear magnetic orderings of $\beta$-MnO$_2$ (FM, A1-AFM, A2-AFM, C-AFM,  and G-AFM) computed using standard DFT (PBEsol functional). The zero of energy corresponds to the Fermi energy (in the case of metallic ground states) or top of the valence bands (in the case of insulating ground states). The intensity of PDOS and total DOS was rescaled to the 6-atoms unit cell for A2-AFM, C-AFM, and G-AFM. The spin-up (upper part) and spin-down (lower part) components of the PDOS are shown on each panel (summed over all atoms of the same type). Mn1 and Mn2 correspond to two Mn atoms with the opposite spin polarizations in the AFM cases.}
\label{fig:pdos_DFT}
\end{figure}

\begin{figure*}[t]
  \includegraphics[width=0.47\linewidth]{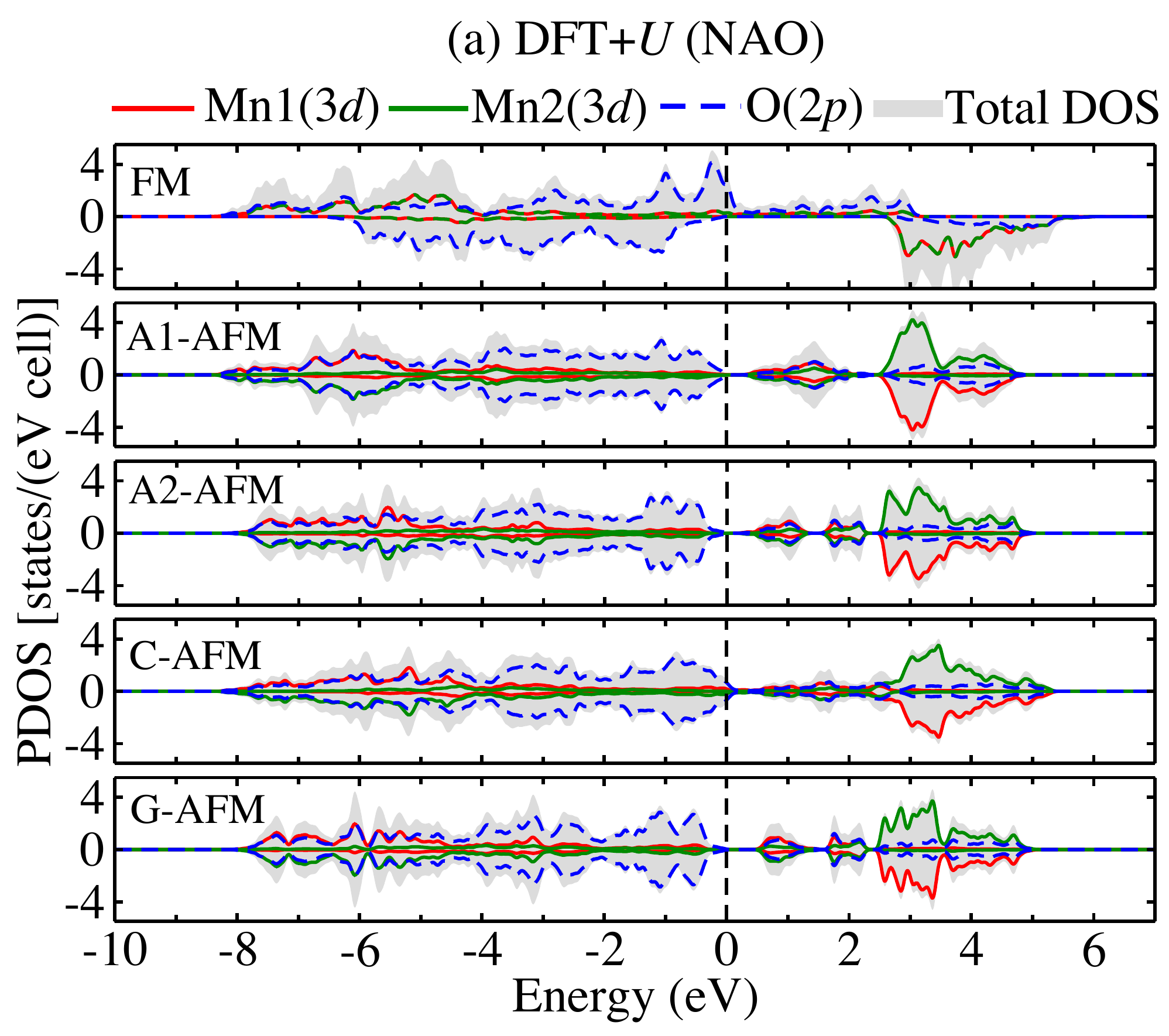}
  \hspace{0.3cm}
  \includegraphics[width=0.47\linewidth]{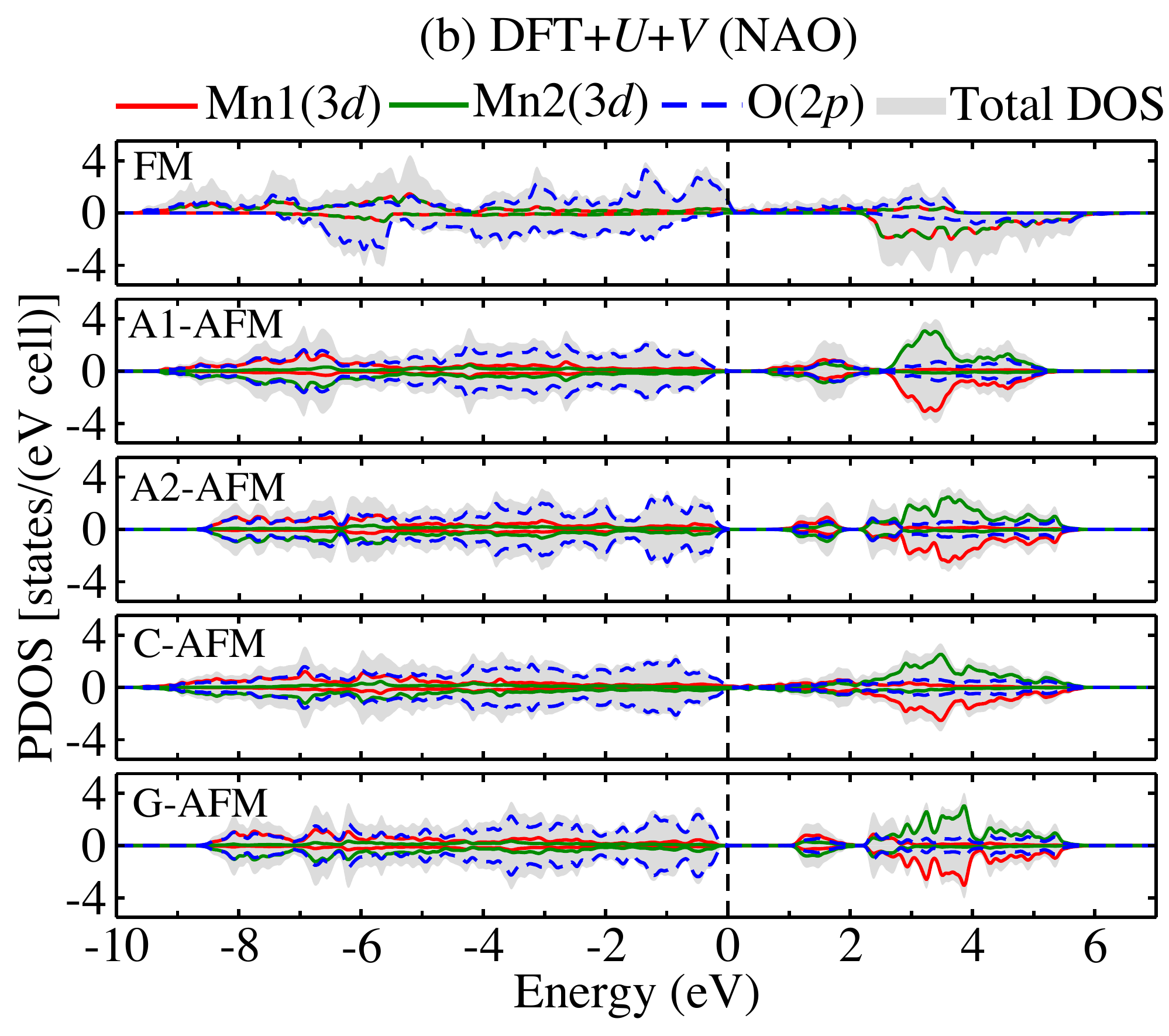}
  \vskip 0.4 cm
  \includegraphics[width=0.47\linewidth]{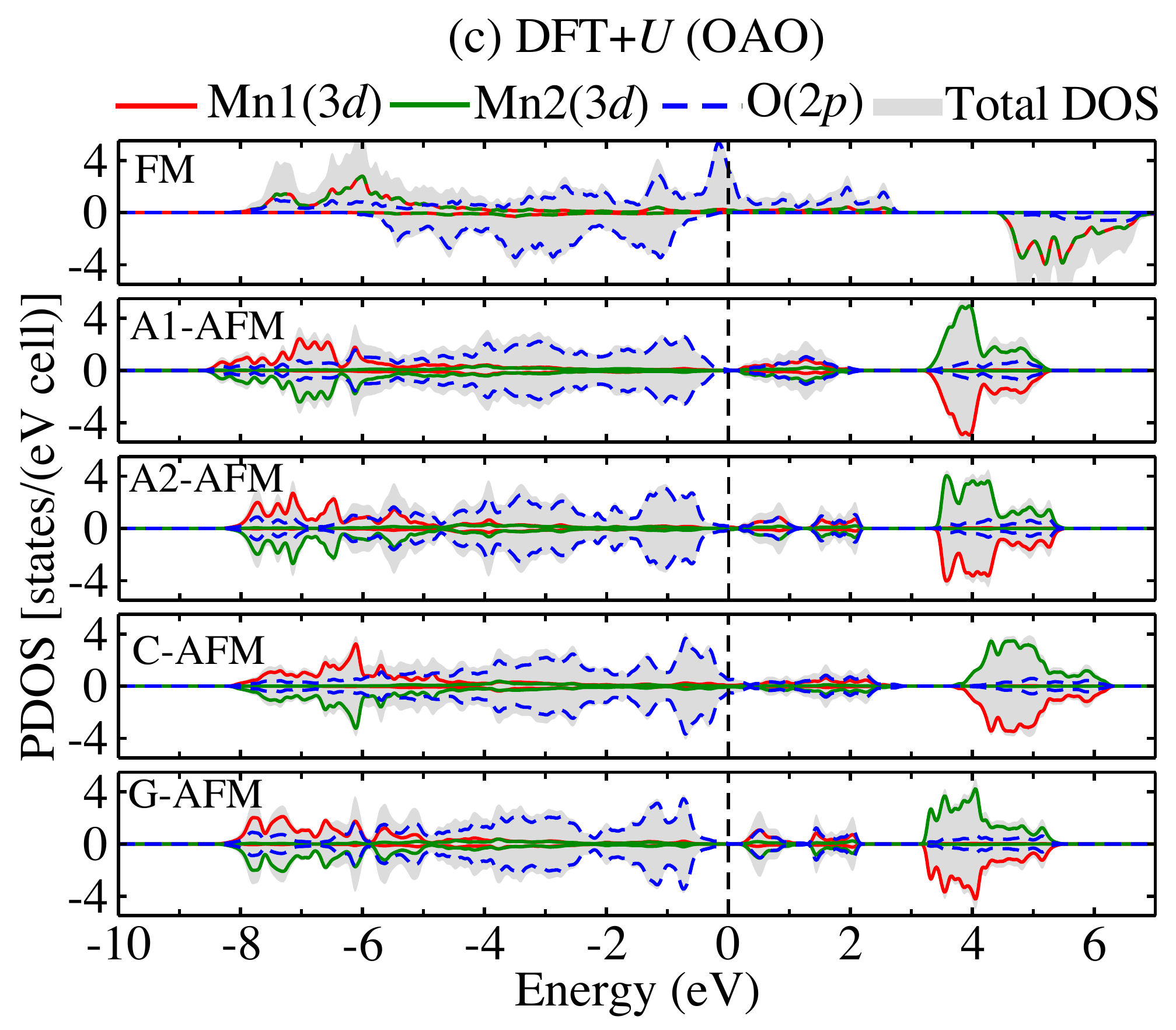}
  \hspace{0.3cm}
  \includegraphics[width=0.47\linewidth]{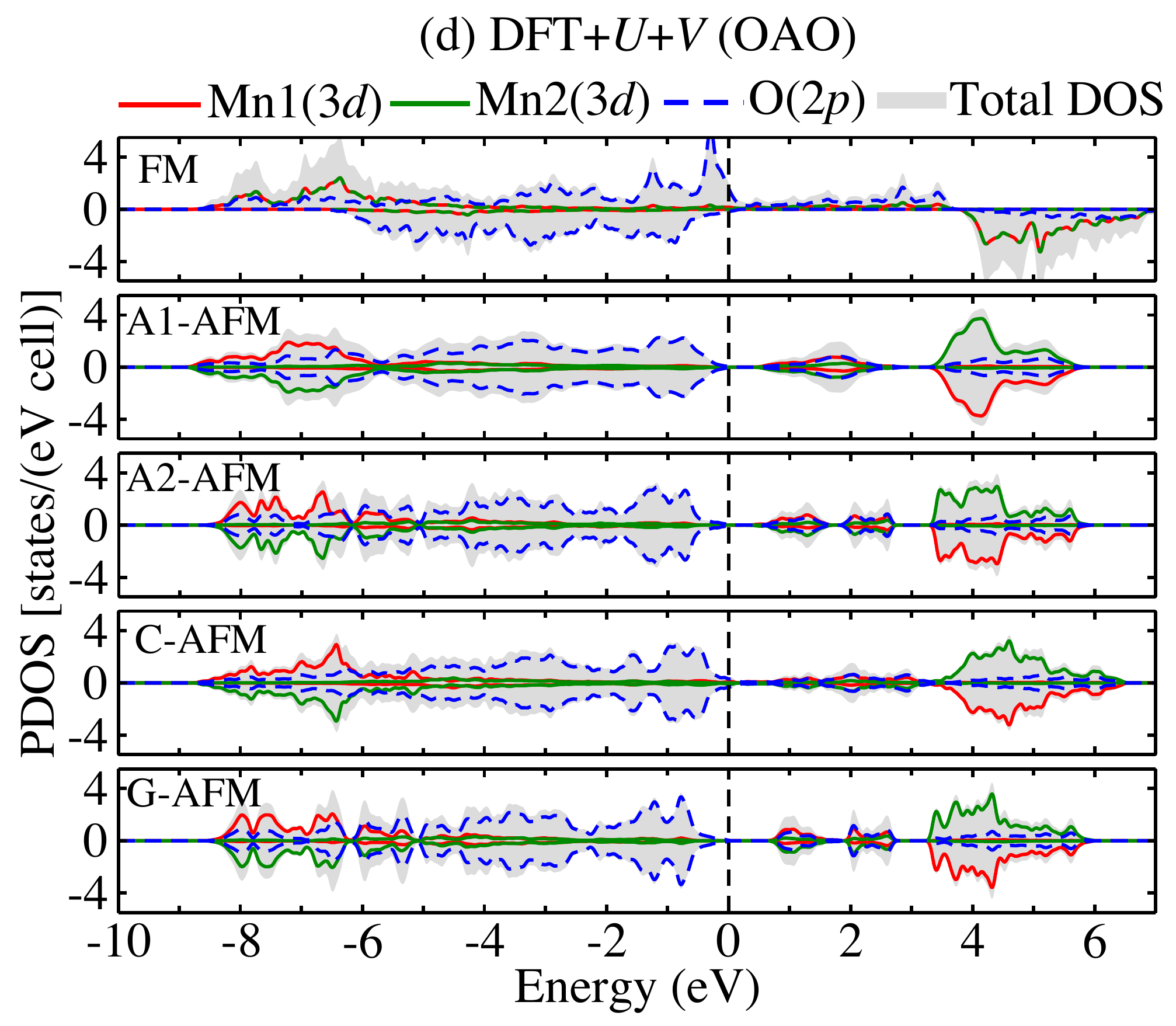}
  \caption{Spin-resolved PDOS and total DOS for five collinear magnetic orderings of $\beta$-MnO$_2$ (FM, A1-AFM, A2-AFM, C-AFM,  and G-AFM) computed at three levels of theory (DFT, DFT+$U$, and DFT+$U$+$V$) using the PBEsol functional. For each case, the Hubbard parameters $U$ and $V$ were computed using DFPT, using two types of Hubbard projectors (NAO and OAO), and they are listed in Table~\ref{tab:Hub_param}. The zero of energy corresponds to the Fermi energy (in the case of metallic ground states) or top of the valence bands (in the case of insulating ground states). The intensity of PDOS and total DOS was rescaled to the 6-atoms unit cell for A2-AFM, C-AFM, and G-AFM. The spin-up (upper part) and spin-down (lower part) components of the PDOS are shown on each panel (summed over all atoms of the same type). Mn1 and Mn2 correspond to two Mn atoms with the opposite spin polarizations in the AFM cases.}
\label{fig:pdos_DFT_Hubbard}
\end{figure*}

The spin-resolved PDOS and total DOS at the DFT level of theory for five collinear magnetic orderings of $\beta$-MnO$_2$ are shown in Fig.~\ref{fig:pdos_DFT}. As was pointed out in Sec.~\ref{sec:Band_gap}, FM, A1-AFM, and C-AFM magnetic orderings correspond to the metallic ground states (zero band gap), while A2-AFM and G-AFM correspond to the insulating ground states. In all cases the conduction bands minimum is predominantly of the Mn($3d$) character while the valence bands maximum has a mixed nature, namely it shows a strong hybridization between the Mn($3d$) and O($2p$) states. Overall, Mn($3d$) states are overdelocalized due to SIE inherent to DFT with approximate xc functionals (such as $\sigma$-GGA which is used in this case), and they spread over the wide energy interval from about $-8$ to 5~eV due to strong hybridization with O($2p$) states. SIE can be fixed by applying the onsite Hubbard $U$ correction to the Mn($3d$) states and the intersite Hubbard $V$ correction between the Mn($3d$) and O($2p$) states; the results are shown in Fig.~\ref{fig:pdos_DFT_Hubbard}.

As can be seen from Figs.~\ref{fig:pdos_DFT_Hubbard}~(a)--(d), the application of the Hubbard $U$ correction to Mn($3d$) states makes them much more localized than in the case of standard DFT, as expected. In particular, the spectral weight of occupied Mn($3d$) states is shifted to lower energies with the maximum intensity in PDOS being in the range from about $-8$ to $-5$~eV, with some variations depending on the type of the Hubbard correction and type of the magnetic ordering. The empty Mn($3d$) states are shifted to higher energies and their intensity is also increased, which is a fingerprint of stronger localization. In contrast to standard DFT (see Fig.~\ref{fig:pdos_DFT}), the top of the valence bands is clearly of the O($2p$) character, while the lowest conduction bands are still of the strong mixed character highlighting the hybridization between Mn($3d$) and O($2p$) states. The higher energy empty states are predominantly of the Mn($3d$) character, with the position of peaks spanning the range from about 2.5 to 6~eV, depending on the magnetic ordering and type of the Hubbard correction. It is worth noting that our DFT+$U$ PDOS computed using NAO or OAO is in qualitatively good agreement with previous DFT+$U$ studies using PAW Hubbard projectors~\cite{noda2016momentum}.

As was discussed previously, our most accurate results (on average) are obtained at the DFT+$U$+$V$ (OAO) level of theory for the A1-AFM magnetic ordering, the PDOS of which is shown in Fig.~\ref{fig:pdos_DFT_Hubbard}~(d). It would be instructive to make a comparison of this PDOS with the valence- and conduction-band spectra measured in the x-ray photoelectron spectroscopy (XPS) and x-ray absorption near-edge structure (XANES) experiments, respectively. As for what concerns the valence-band XPS measurements, we are aware only of Ref.~\cite{Audi:2002}; however, the resolution of these XPS spectra is not high enough to make a precise comparison with our PDOS. Also, we are not aware of any XANES measurements for $\beta$-MnO$_2$, hence we cannot verify the accuracy of empty-states PDOS computed in this work. However, we want to point out that in the previous study of the XANES spectra of pristine and Ni-substituted LaFeO$_3$ we found that DFT+$U$ and DFT+$U$+$V$ with OAO projectors give results that are in fair agreement with the experimental XANES spectra~\cite{Timrov:2020c}. Therefore, high-resolution XPS and XANES experiments on $\beta$-MnO$_2$ are called for.

Finally, for the sake of completeness, in the Appendix we present the spin-resolved PDOS of the $t_{2g}$ and $e_g$ components of the Mn($3d$) orbitals for the A1-AFM ordering computed using DFT, DFT+$U$ (OAO) and DFT+$U$+$V$ (OAO). Similar trends are found also for other AFM orderings and when using the NAO Hubbard projectors.

\section{Conclusions}
\label{sec:Conclusions}

We have presented a fully first-principles investigation of the structural, electronic, and magnetic properties of five collinear magnetic orderings of $\beta$-MnO$_2$ (FM, A1-AFM, A2-AFM, C-AFM, G-AFM). A Hubbard-corrected DFT approach was used with onsite $U$ and intersite $V$ Hubbard parameters computed using density-functional perturbation theory~\cite{timrov2018hubbard, Timrov:2021}, with two types of Hubbard projectors based on nonorthogonalized and orthogonalized atomic orbitals. 

We have found that there are sizable variations in the predicted results depending on the type of the spin configuration, on whether only $U$ or both $U$ and $V$ corrections are applied, and on the type of the Hubbard projectors. The most accurate results on average are found using DFT+$U$+$V$ with OAO projectors for the A1-AFM magnetic ordering. At this level of theory, A1-AFM is found to be the most energetically favorable compared to FM and all other AFM orderings. This finding suggests that the commonly used collinear A1-AFM ordering as an approximation to the true noncollinear helical ordering is justified when the intersite Hubbard interactions are taken into account and when Hubbard projectors are carefully chosen. Indeed, we have  highlighted the crucial role played by the intersite $V$ correction to describe accurately strong hybridizations between Mn($3d$) and O($2p$) states in $\beta$-MnO$_2$. 

Within DFT+$U$+$V$ (OAO), a semiconducting ground state was obtained for A1-AFM with the band gap of 0.32~eV which is in good agreement with the experimental values of $0.26-0.28$~eV~\cite{Chevillot:1959, yu2008electronic, Islam:2005, druilhe1967electron}, and the magnetic moment on Mn atoms is predicted to be 2.74~$\mu_\mathrm{B}$ which is in a satisfactory agreement with the experimental value of 2.35~$\mu_\mathrm{B}$~\cite{regulski2003incommensurate}. The remaining discrepancies between our best computed results and the experimental ones can be attributed to the fact that A1-AFM is just an approximation to the noncollinear helical magnetic ordering, and perhaps the inclusion of Hund's $J$ could also bring our theoretical predictions even closer to the experimental ones. Therefore, the generalization of DFT+$U$+$V$ to include $J$ and the investigation of the complex noncollinear helical magnetic ordering are topics of future studies on $\beta$-MnO$_2$. Finally, it would be interesting to further examine the accuracy of DFT+$U$+$V$ with OAO Hubbard projectors for other properties of $\beta$-MnO$_2$ such as phonons, thermoelectric and thermochemical properties, formation energy of oxygen vacancies, to name a few, as well as to analyze the effective hoppings and magnetic interaction parameters.

The data used to produce the results of this paper are available in the Materials Cloud Archive~\cite{MaterialsCloudArchive2021}.

\begin{figure}[t!]
\begin{center}
  \includegraphics[angle=-90,width=0.47\textwidth]{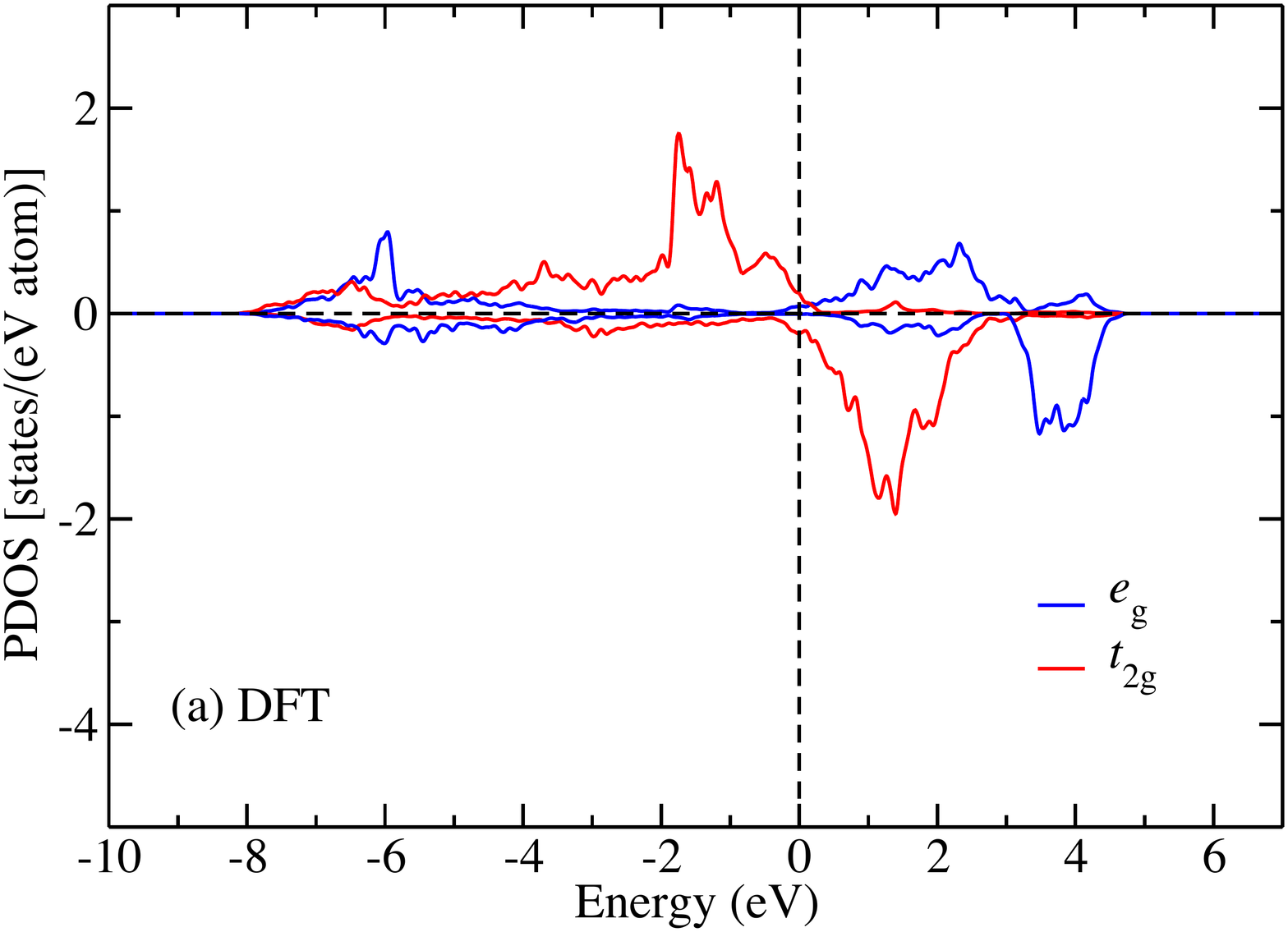}
  \vskip 0.3 cm
  \includegraphics[angle=-90,width=0.47\textwidth]{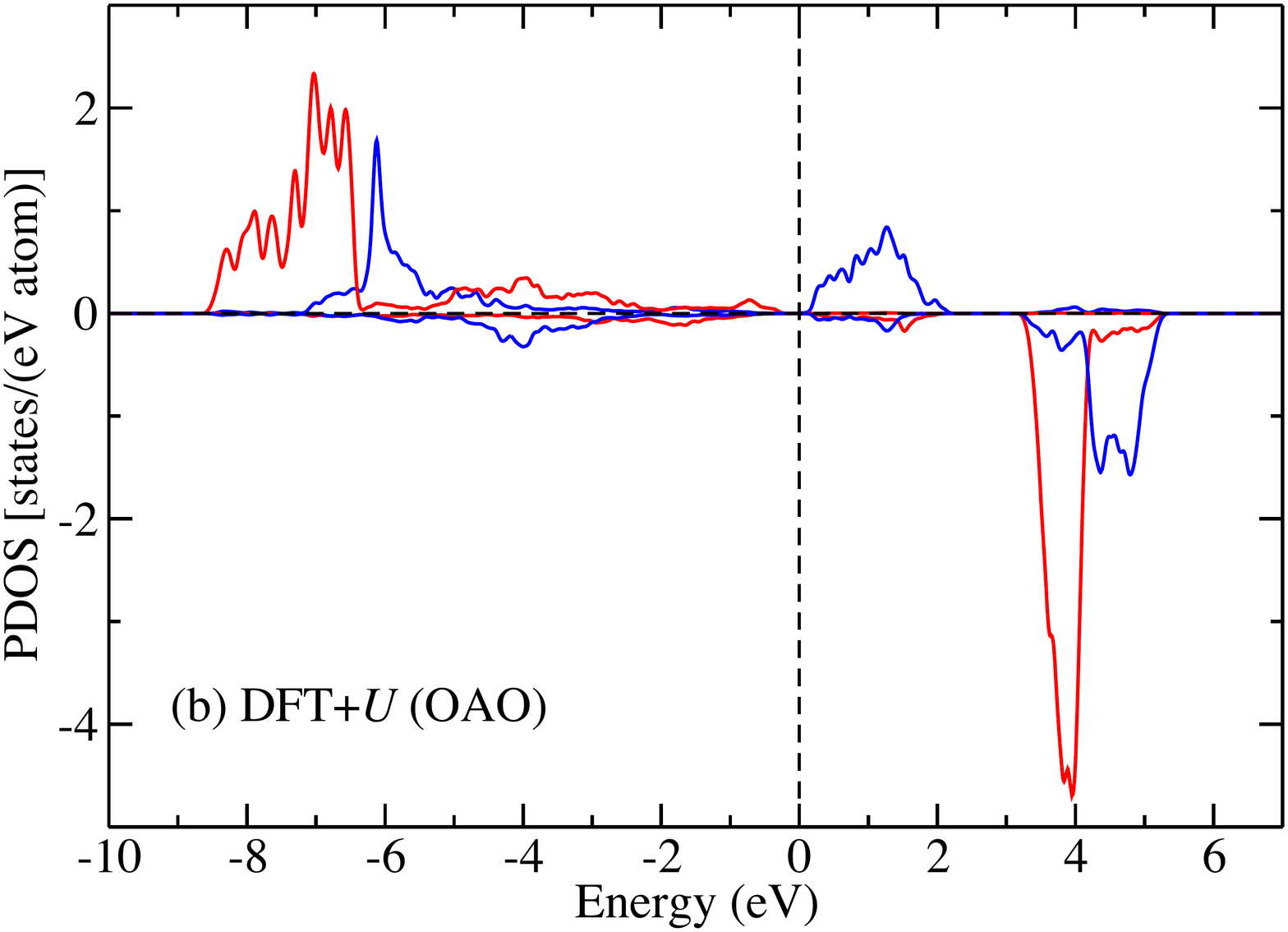}
  \vskip 0.3 cm
  \includegraphics[angle=-90,width=0.47\textwidth]{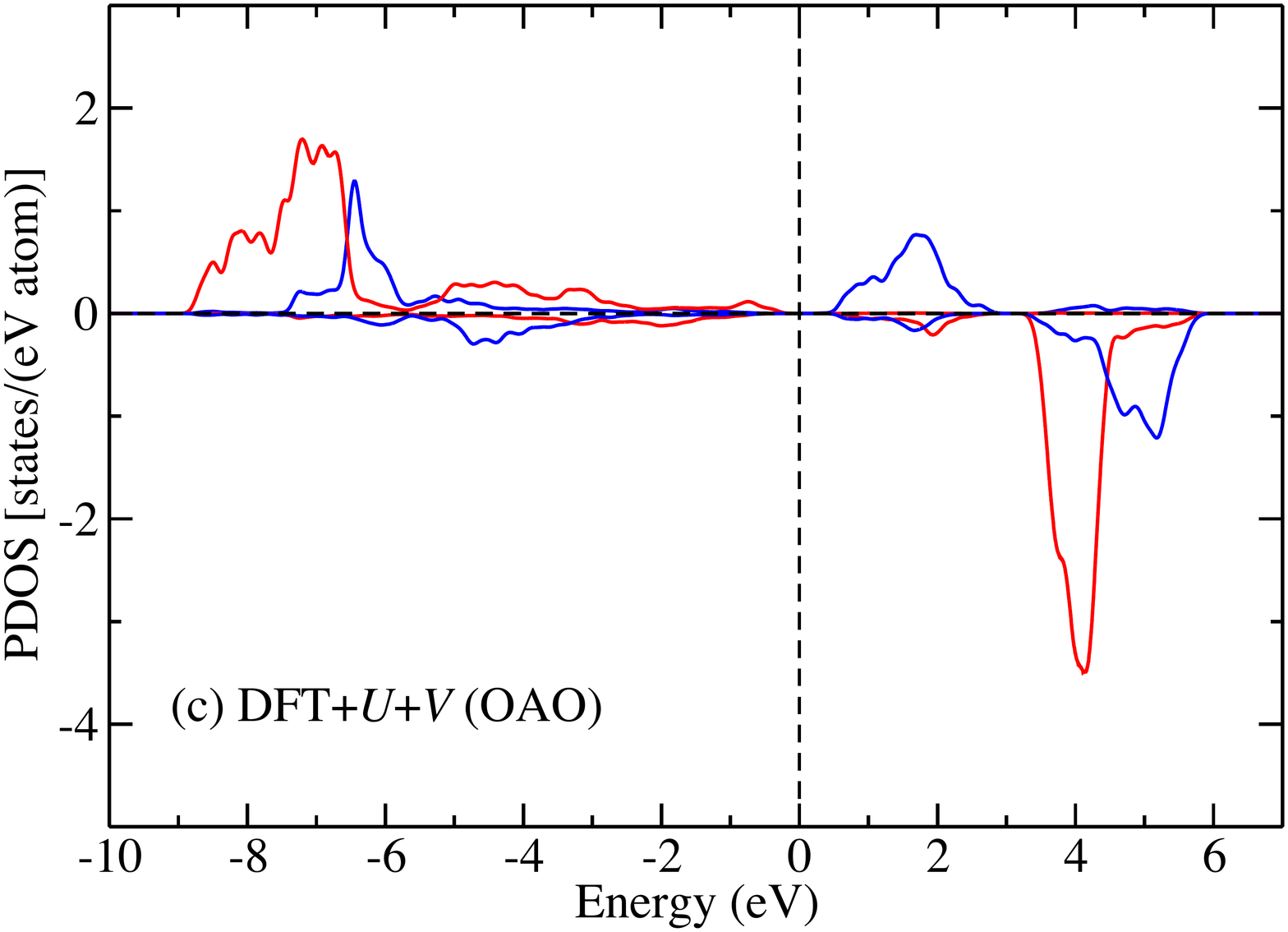}
  \caption{Spin-resolved PDOS for the $t_{2g}$ and $e_g$ states of one Mn atom of $\beta$-MnO$_2$ with the A1-AFM magnetic ordering computed at three levels of theory (DFT, DFT+$U$, and DFT+$U$+$V$) using the PBEsol functional. OAO Hubbard projectors are used. The upper and lower panels on each figure correspond to the spin-up and spin-down channels, respectively. The zero of energy corresponds to the Fermi energy (in the DFT case) or top of the valence bands (in the case of DFT+$U$ and DFT+$U$+$V$).}
\label{fig:app}
\end{center}
\end{figure}

\section*{Acknowledgments}

We thank Francesco Aquilante for fruitful discussions. R.M. acknowledges support by the IIT Mandi for the HTRA fellowship. I.T. and N.M. acknowledge support by the Swiss National Science Foundation (SNSF), through Grant No.~200021-179138, and its National Centre of Competence in Research (NCCR) MARVEL. Computer time was provided by the Holland Computing Centre (University of Nebraska) and by the Swiss National Supercomputing Centre (CSCS) under Project No.~s836 and No.~s1073.


\appendix

\section*{Appendix: Spin-resolved PDOS of the Mn($3d$) orbitals}
\label{app:PDOS_Mn}

In this Appendix we present the spin-resolved PDOS of the $t_{2g}$ and $e_g$ components of the Mn($3d$) orbitals for the A1-AFM ordering computed using DFT, DFT+$U$ (OAO) and DFT+$U$+$V$ (OAO). 

The calculation of the PDOS for the $t_{2g}$ and $e_g$ orbitals was done by diagonalizing the onsite occupation matrix [see Eq.~\eqref{eq:occ_matrix_0}, with $I=J$], then using its eigenvectors to rotate the OAO (that are defined in the global frame), and finally by projecting the KS wave functions onto these rotated orbitals and computing 
\begin{equation}
\mathcal{I}^{I,\sigma}_{m}(\varepsilon) = \sum_{v,\mathbf{k}} f_{v,\mathbf{k}}^\sigma |\langle \psi^\sigma_{v,\mathbf{k}} | \tilde{\phi}^I_{m,\mathrm{rot}} \rangle|^2 \delta(\varepsilon - \varepsilon_{v,\mathbf{k}}) \,,
\label{eq:PDOS}
\end{equation}
where $\varepsilon_{v,\mathbf{k}}$ are the KS energies, $\tilde{\phi}^I_{m,\mathrm{rot}}$ are the rotated OAO, and all other quantities are defined in Sec.~\ref{sec:methods}. The rotation of the orbitals is needed in order to obtain the PDOS in a local frame and not in the global frame -- this allows us to assign states to the $t_{2g}$ or $e_g$ character (we adopt the approximate but more insightful $O_h$-group nomenclature, although it is understood that locally the point group symmetry is reduced to $D_{2h}$). We have implemented the calculation of PDOS with the rotated orbitals in \QE{} and will make it publicly available in the next official release.

The PDOS computed using Eq.~\eqref{eq:PDOS} is shown in Fig.~\ref{fig:app}. It is seen that there are large differences (between DFT and Hubbard-corrected DFT results) in the shape and position of the $t_{2g}$ and $e_g$ states in both spin channels. In particular, in DFT the $t_{2g}(\uparrow)$ and $t_{2g}(\downarrow)$ states are very close to the Fermi level and they contribute to the DOS at the Fermi level (to recall, the ground state is metallic in DFT for A1-AFM); moreover, $t_{2g}(\uparrow)$ are essentially fully occupied while $t_{2g}(\downarrow)$ are mainly empty, as can also be see from the occupation numbers in Table~\ref{tab:Hub_occupations}. The $e_{g}(\uparrow)$ and $e_{g}(\downarrow)$ states in DFT are both partially occupied (see Table~\ref{tab:Hub_occupations}). The inclusion of the Hubbard corrections ($U$ and $V$) dramatically changes the PDOS: the $t_{2g}(\uparrow)$ states are pushed downwards in energy, while $t_{2g}(\downarrow)$ are pushed upwards and they become sharper [especially $t_{2g}(\downarrow)$], while the $e_{g}(\uparrow)$ and $e_{g}(\downarrow)$ states also experince significant modifications as can be seen in Fig.~\ref{fig:app}. An interesting difference between the DFT+$U$ and DFT+$U$+$V$ PDOS is that the peak associated with $t_{2g}(\downarrow)$ has larger intensity in DFT+$U$ than in DFT+$U$+$V$ - this is so because $U$ in DFT+$U$ favors the localization of these states while $V$ in DFT+$U$+$V$ counterbalances  the effect of $U$ by favoring the hybridization with O($2p$) states that reduces the intensity of the $t_{2g}(\downarrow)$ peak. The reduction of the peak intensity in DFT+$U$+$V$ is also observed for other states in Fig.~\ref{fig:app}. As for what concerns the occupation numbers, DFT+$U$ and DFT+$U$+$V$ give similar values as can be seen in Table~\ref{tab:Hub_occupations}, and they differ considerably from the DFT occupations. In particular, we can see that in the Hubbard-corrected DFT the $t_{2g}(\uparrow)$ states are fully occupied ($\sim 1.0$), $t_{2g}(\downarrow)$ are almost fully empty, while $e_{g}(\uparrow)$ and $e_{g}(\downarrow)$ are partially occupied.

\begin{table}[t]
\renewcommand{\arraystretch}{1.3}
\centering
\begin{tabular}{cccc}
\hline\hline
\parbox{2.5cm}{Orbital character} & \parbox{1.5cm}{DFT} &  \parbox{1.5cm}{DFT+$U$} & \parbox{1.5cm}{DFT+$U$+$V$}  \\ \hline
\multirow{3}{*}{\begin{tabular}[c]{@{}c@{}}\parbox{1cm}{$t_{2g}(\uparrow)$}\end{tabular}}  
                    &   0.994             &          0.993           &      0.994     \\
                    &   0.969             &          0.996           &      0.996     \\
                    &   0.969             &          0.996           &      0.996     \\
\multirow{2}{*}{\begin{tabular}[c]{@{}c@{}}\parbox{1cm}{$e_{g}(\uparrow)$}\end{tabular}}  
                    &   0.455             &          0.571           &      0.534     \\
                    &   0.470             &          0.586           &      0.547     \\ \hline
\multirow{3}{*}{\begin{tabular}[c]{@{}c@{}}\parbox{1cm}{$t_{2g}(\downarrow)$}\end{tabular}}  
                    &   0.164             &          0.054           &      0.062     \\
                    &   0.224             &          0.101           &      0.120     \\
                    &   0.269             &          0.111           &      0.141     \\
\multirow{2}{*}{\begin{tabular}[c]{@{}c@{}}\parbox{1cm}{$e_{g}(\downarrow)$}\end{tabular}}  
                    &   0.342             &          0.265           &      0.291     \\
                    &   0.355             &          0.278           &      0.304     \\
\hline\hline
\end{tabular}%
\caption{Eigenvalues of the occupation matrix, Eq.~\eqref{eq:occ_matrix_0} (with $I=J$), for one Mn atom. Spin-up and spin-down components of the $t_{2g}$ and $e_g$ states are indicated with up arrow and down arrow, respectively. The DFT+$U$ and DFT+$U$+$V$ results were obtained using the OAO Hubbard projectors.}
\label{tab:Hub_occupations}
\end{table}

Finally, it is instructive to discuss the splitting between the $t_{2g}(\downarrow)$ and $e_{g}(\downarrow)$ states. In DFT, there is clearly a gap between these states (i.e., crystal field splitting), while in DFT+$U$ and DFT+$U$+$V$ the separation of these states is reduced so strongly that they start to overlap. This effect can be explained based on the arguments of the crystal field theory. The splitting between the $d$ states depends on the distance between the ligands (O) and a metal ion (Mn): the smaller the Mn-O distance the larger the splitting of $d$ states. As can be seen from Fig.~\ref{fig:crystal_structure} and Fig.~\ref{fig:Interatomic_dist}~(a) (second panel, A1-AFM), the Mn-O distance in DFT is smaller than in DFT+$U$ and DFT+$U$+$V$, and as a result the splitting between the $t_{2g}(\downarrow)$ and $e_{g}(\downarrow)$ states is larger in DFT than in DFT+$U$ and DFT+$U$+$V$. It is worth pointing out that a similar result was obtained in Ref.~\cite{tompsett2012importance}.


%

\end{document}